\documentclass[12pt,fleqn]{article} \textheight=9in
\newcommand{\cf}{{\em cf. }}
\newcommand{\eg}{{\em e.g.}}
\newcommand{\ra}{\rightarrow}
\newcommand{\bra}{\langle} \newcommand{\ket}{\rangle}
\newcommand{\be}{\begin{equation}}
\newcommand{\ee}{\end{equation}}
\newcommand{\bea}{\begin{eqnarray}}
\newcommand{\eea}{\end{eqnarray}}
\newcommand{\eps}{\epsilon}
\newcommand{\e}{\mbox{e}}

\newcommand{\E}{\mbox{e}}
\newcommand{\ffi}{\varphi}
\newcommand{\sign}{\mbox{\rm sign}}
\newcommand{\ep}{\hfill \fbox{}}
\newcommand{\ode}{{\cal O}}
\newcommand{\A}{\cal A}
\newcommand{\B}{\cal B}

\newtheorem{claim}{Claim}[section]
\newtheorem{thm}[claim]{Theorem}
\newtheorem{prop}[claim]{Proposition}
\newtheorem{lem}[claim]{Lemma}
\newtheorem{cor}[claim]{Corollary}

\def\R{\mathbb{R}}

\def\Z{\mathbb{Z}}
\def\N{\mathbb{N}}
\def\un{\mathbb{I}}
\usepackage{amssymb}

\begin{document}

\title{Avoided crossings in mesoscopic systems:
\\  electron propagation on a non-uniform magnetic cylinder}
\author{P. Exner$^{a,b}$ and A. Joye$^{c}$}
\date{}
\maketitle
\begin{quote}
{\small \em a) Nuclear Physics Institute, Academy of Sciences,
CZ--25068 \v Re\v z near Prague \\ b) Doppler Institute, Czech
Technical Univ., B\v rehov{\'a} 7, CZ-11519 Prague \\ c) Institut
Fourier, Universit\'{e} de Grenoble 1, F-38402 Saint-Martin
d'Heres
\\ \rm \phantom{e)x}exner@ujf.cas.cz, joye@ujf-grenoble.fr}
\vspace{8mm}

\noindent {\small We consider an electron constrained to move on a
surface with revolution symmetry in the presence of a constant
magnetic field $B$ parallel to the surface axis. Depending on $B$
and the surface geometry the transverse part of the spectrum
typically exhibits many crossings which change to avoided
crossings if a weak symmetry breaking interaction is introduced.
We study the effect of such perturbations on the quantum
propagation. This problem admits a natural reformulation to which
tools from molecular dynamics can be applied. In turn, this leads
to the study of a perturbation theory for the time dependent
Born-Oppenheimer approximation.}
\end{quote}

\vspace{3mm}
\section{Introduction}

Recent advances in experimental physics have made it possible to
produce two dimensional conducting surfaces of mesoscopic size. In
such devices, the mean free path often exceeds the system size so
the electron motion is ballistic and quantum coherence effects
play a crucial role. This gives a motivation to strive for a
complete understanding of the quantum mechanics of corresponding
processes. In particular, conducting carbon ``nanotubes'' which
are more or less uniform cylinders, belong to the family of
surfaces that are nowadays experimentally within reach \cite{i}.
Since their discovery, lots of studies have been devoted to the
elucidation of the spectral and transport properties of such
devices, in a variety of situations and approximations -- see,
e.g. \cite{a, lr, wfas} and references therein. Nanotubes of
different types can be combined, and also coupled to other carbon
structures such as fullerene molecules \cite{bamboo},  producing a
variety of cylindrical surfaces .

In this paper we study a quantum propagation in an ``imperfect
nanotube'' subject to a constant magnetic field parallel to the
tube axis within a simple model. Our model assumption is that a
single electron is confined to a surface of revolution with slow
variation of the radius along the revolution axis. Moreover, we
assume that the rotational symmetry is weakly violated, either by
an impurity or by an external field. In other words, the used
idealization amounts to neglecting the atomic structure of the
tube as well as the interaction between the electrons, but taking
into account the gross shape of the device. Our aim is to study
the propagation of the electron along such an imperfect nanotube
in the homogenous magnetic field by means of the time-dependent
Schr\"odinger equation, starting with an initially localized wave
packet, and paying a particular attention to the transitions
between angular levels caused by the symmetry breaking
perturbation.

To understand the peculiarities of this quantum mechanical
problem, it is useful to review briefly its classical counterpart;
this is done in Section~2. The first question in the quantum case
is the meaning of the fact that the electron is confined to a
surface. The most natural approach, to our opinion, treats the
surface as a limiting situation of a thin hard-wall layer. This
idea goes back to \cite{dC1, dC2, To} and requires a
renormalization in which the transverse contribution to the energy
-- blowing-up in the limit -- is removed. One gets in this way an
additional curvature-dependent term, in general attractive, to the
potential. For the sake of completeness recall that there has been
another recent work treating particle motion on revolution
surfaces --- see \cite{Al, Ar, MV} and references therein. The
last two papers aim at solvable models of compact surfaces
(neglecting the curvature-dependent term), while \cite{Al} treats
the Schr\"odinger and wave equations on noncompact cylindrical
surfaces without a magnetic field from the PDE point of view.

Having thus found the Hamiltonian of our quantum system, we can
analyze its spectral properties. When the rotational symmetry is
preserved, we can perform (using a suitable gauge) the
partial-wave decomposition. We can compute the angular part of the
spectrum which depends on the actual cylinder radius varying along
the tube axis. This brings to mind analogy with the molecular
dynamics in which each angular state corresponds to an ``electronic''
level and the longitudinal coordinate measured at the axis
corresponds to the one-dimensional ''nuclear'' configuration
coordinate. Furthermore, when the rotational symmetry is broken by
a perturbation, the above analogy remains valid and we may invoke
the time-dependent Born-Oppenheimer approximation to describe the
propagation along coupled angular levels according to
\cite{master, hj, hj2}. Recall that the theory in molecular
systems involves a small parameter which is given by the mass
ratio between ``electrons'' and ''nuclei''. In our mesoscopic system, it
is replaced by the parameter $\epsilon$ defined as the inverse of
the lenghtscale over which the variation of the radius of the
nanotube takes place. Note, however, that we cannot directly apply
the theory of \cite{master, hj, hj2} in our perturbative context
and a modification is needed as we shall explain below.

The result of the analysis presented below gives a complete and
rigorous description to the leading order of the wave function
when the dynamics makes the electron go through a region where a
perturbation couples angular levels. The basic picture is as
follows. As long as the perturbed angular levels along the
trajectory remain well isolated, the components of the wave
function referring to the corresponding eigenstates are unchanged,
to the leading order. When the unperturbed angular levels display
a crossing or an avoided crossing, transitions between the
perturbed angular eigenstates may become non-negligible as in the
mentioned molecular analogy -- see \cite{hj}. We are going to
consider precisely the situation where the transition amplitudes
are of order one, under perturbations of order $\sqrt{\epsilon}$.
In such cases, an initial wave function having a nonzero component
in a single angular eigenspace before the (avoided) crossing
splits into the corresponding angular eigenstates according to the
Landau-Zener formula, again to the leading order.

We have said already that despite being based on the paper
\cite{hj}, our description is not a direct application of the
molecular time-dependent Born-Oppenheimer approximation. Indeed,
in the rigorous derivations of this approximation, the ``electronic''
spectrum and the eigenstates, i.e. the spectrum at fixed
coordinate along the rotation axis in our setting, are taken as
given data, and the approximate solution to the molecular
Schr\"odinger equation is constructed from this information -- see
\cite{master, hj, hj2}. In our situation, by hypothesis, we only
have access to that spectrum in a perturbative sense, and thus we
need to develop a perturbative version of the time-dependent
Born-Oppenheimer approximation that only requires knowledge of the
leading terms of the Rayleigh-Schr\"odinger perturbation series.
{This is done in Section 5, where }
the main technical result of the paper is stated in Theorem
\ref{57}. We believe that it is of an independent interest.

Precise statements of our results require a certain amount of
notation and are therefore given below in Proposition \ref{aprox} 
and Theorems \ref{57} .


\setcounter{equation}{0}
\section{Classical Mechanics}

Let us start by describing the classical dynamics of the system.
We consider a particle of mass $m$ and charge $e$ constrained to
move on a smooth surface $S$ with revolution symmetry around the
axis $OX$ in a homogeneous magnetic field ${\bf B}=B e_x$, $B\geq
0$, parallel to this axis.

Using cylindrical coordinates, the surface is characterized by the
smooth positive real valued function $\R\ni x\mapsto R(x)\in
\R_+^*$ such that
\be
  \left.\matrix{x &\!= \!& x \cr y &\!= \!& R(x)\cos(\theta)\cr z &\!= \!& R(x)\sin(\theta)}\right\}
\ee
where $(x,\theta)\in \R\times S^1$. The squared length element on
$S$ is $ds^2 = (1+R'(x)^2) dx^2 + R(x)^2 d\theta^2$, so the
corresponding metric tensor $g_{ij}(x,\theta)$  is given by
\be \label{g_ij}
  g_{ij}(x,\theta)=\pmatrix{1+{R'(x)}^2 & 0\cr 0& R^2(x)}.
\ee
Using the circular gauge, we express the vector potential at the
surface as
\be
  {\bf A(r)}=\frac{1}{2}{\bf B\wedge r}=\frac{R(x)B}{2}\pmatrix{0\cr
  -\sin(\theta)\cr \cos(\theta)}
\ee
This makes it possible to compute the Lagrangian function of the
system
\bea
  L({\bf r},\dot{\bf r}) &\!= \!& \frac{1}{2}m\dot{\bf r}^2+e\dot{\bf r}
  {\bf A(r)}\nonumber \\
   &\!= \!& \frac{1}{2}m\left(\dot{x}^2(1+{R'(x)}^2)+R^2(x)\dot{\theta}^2\right)
  +\frac{eBR^2(x)\dot{\theta}}{2}\,.
\eea
The system is integrable: we find that the momentum $p_{\theta}=
\frac{\partial L}{\partial \dot{\theta}}$ and the kinetic energy
$T$ are two constants of motion,
\bea\label{ptheta}
 p_{\theta} &\!= \!& mR^2(x)\dot{\theta}+\frac{eBR^2(x)}{2}\,,\\ \label{kin}
 T &\!= \!& \frac{1}{2}m\left(\dot{x}^2(1+{R'(x)}^2)+R^2(x)\dot{\theta}^2\right).
\eea
Using (\ref{ptheta}) to express $\dot{\theta}$ as a function
of $x$ in (\ref{kin}), we deduce
\bea
  T &\!= \!& \frac{1}{2}m\dot{x}^2(1+{R'(x)}^2)+\frac{1}{2m}
  \left(\frac{p_{\theta}^2}{R^2(x)}-p_{\theta}eB+
  \frac{e^2B^2}{4}R^2(x)\right)\nonumber\\ \label{effpot}
   &\!= \!& \frac{1}{2}m\dot{x}^2(1+{R'(x)}^2)+V(R(x)).
\eea
The effective potential  $\R_+^*\ni R\mapsto V(R)\in \R_+$ admits
a unique minimum at $R_0$ such that
\be
  R_0=\sqrt{\frac{2|p_{\theta}|}{|e|B}}\,\,\,\mbox{ and }
  \,\,\, V(R_0)=\left\{\matrix{0 &\mbox{if}& ep_{\theta}\geq 0\cr
  \frac{ |ep_{\theta}| B}{m} &\mbox{if}& ep_{\theta}< 0}\right.
\ee
Note that if $p_{\theta}=0$, the potential $V(R)$ is harmonic on
$\R_+^*$. From these considerations we deduce easily, in
particular, that particle motions in the simplest case
characterized by $\dot{x}(t)=0$ correspond either to
$(x(t),\theta(t))=(x_0,\theta_0)$ for any initial conditions
$(x_0,\theta_0)$, or to $(x(t),\theta(t))=
(x_0,\theta_0-\frac{eB}{m} t)$, where $\frac{|e|B}{m}=:\omega_c$
is the cyclotronic frequency, for any initial conditions
$(x_0,\theta_0)$, or finally to $(x(t),\theta(t))=
(x_0,\theta_0+\omega t)$, where $\omega$ is any constant, for
initial conditions  $(x_0,\theta_0)$ such that $R'(x_0)=0$. In
case that $R'(x_0)\neq 0$, the first two motions are stable,
whereas in the last one the stability depends on the local
properties of $R$ around $x_0$. In a similar way one can treat the
general case with $\dot{x}(t)\neq 0$. The motion is governed by
the effective potential determined by the shape of $S$, and the
potential minima correspond to the points where the angular motion
has the cyclotronic frequency.

Furthermore, notice that the addition of a supplementary exterior
potential $W$, depending on $x$ only, does not affect the
functional dependence of $ p_{\theta}$ and its value remains
independent of time. It is just the second constant of motion
which is changed at that in the sense that the total energy
$E=T+W$ is now constant.

Finally, let us also give the corresponding Hamiltonian function
of the system for a future purpose. With $ p_x= \frac{\partial
L}{\partial \dot{x}}$ we compute
\be
 H(x,\theta,p_x, p_{\theta})=\left( \frac{p_x^2}{2m(1+{R'(x)}^2)}
 +\frac{1}{2mR^2(x)}
 \left( p_{\theta}-\frac{eBR^2(x)}{2}\right)^2\right). \label{clH}
\ee
In the sequel we shall consider our charged particle to be an
electron, $e=-|e|<0$, and use the rational units in which
$|e|=m=1$ as well as $\hbar=c=1$.


\setcounter{equation}{0}
\section{Quantum Mechanics}

Consider now the same system within quantum mechanics. For the
purpose of this section, the function $\,R: \R\to\R_+\,$ defining
the surface  $S$ is supposed to strictly positive and $\,C^3$ smooth;
later we shall impose stronger requirements.

The state Hilbert space of such a system is thus $L^2(S)$. To
construct the Hamiltonian, however, it is not sufficient to
replace the classical variables in (\ref{clH}) by the
corresponding canonical operators. The most natural way of
quantization consists of taking a particle confined within a
cylindrical layer built over $S$ and squeezing its thickness to
zero --- \cf\cite{dC1, dC2, To}. One has to renormalize the energy
in the limit, of course, subtracting the blowing-up part
corresponding to the transverse motion.

In the absence of the magnetic field, one arrives in this way to
the Hamiltonian which equals to a sum of the respective
Laplace-Beltrami operator (times $1/2$ in our units) and the
curvature-induced potential $ V(x) = -\,{1\over 8}
\left(\varrho_1(x)^{-1}\! - \!\varrho_2(x)^{-1}\right)$, where
$\,\varrho_j(x),\: j=1,2\,$, are the principal curvature radii at
the given point. The second part is of a purely quantum nature and
has no classical counterpart. In the present case the locally
elliptical intersection of $S$ with the normal plane has the
radius $\varrho_1(x) = R(x)\,$, while for the intersection with
the axial plane we find
\be
 \varrho_2(x) = -\, {(1+R'(x)^2)^{3/2} \over R''(x)}\;;
\ee
the signs of $\varrho_1,\, \varrho_2$ coincide if both the
osculation radii point the same side of the surface. Consequently,
the curvature-induced potential equals
\be \label{curvature potential}
 V(x) = -\, {1\over 8R(x)^2}\,
 \left(1+ {R(x)R''(x) \over (1+R'(x)^2)^{3/2}}\right)\,.
\ee
To express the kinetic (Laplace-Beltrami) part, $-\,{1\over 2}\,
|g|^{-1/2} \partial_i |g|^{1/2} g^{ij} \partial_j$, we use
(\ref{g_ij}) and the corresponding contravariant tensor on $S$,
\be\label{g^ij}
 (g^{ij}(x)) = \left( \begin{array}{cc}
 (1+R'(x)^2)^{-1} & 0 \\ 0 & R(x)^{-2} \end{array} \right)\,.
\ee
The Hamiltonian in the presence of the magnetic field is then
obtained by replacing the angular momentum operator $p_{\theta}=
-i\partial_{\theta}$ by $p_{\theta}-A(x)R(x)\,$ where $A(x):=
A_\theta(\bf r)$; it acts as
\bea \label{Hamiltonian1}
 H &\!= \!&  -\, {1\over 2R(x)
 \sqrt{1+R'(x)^2}}\, \partial_x\, {R(x) \over \sqrt{1+R'(x)^2}}\,
 \partial_x\,+\, {1\over 2R(x)^2}\, \left( -i\partial_{\theta}\,+\,
 {BR(x)^2\over 2} \right)^2 \nonumber \\ && \nonumber \\ && -\,
 {1\over 8R(x)^2}\, \left( 1+ {R(x)R''(x) \over (1+R'(x)^2)^{3/2}}
 \right)^2
\eea
on an appropriate domain in $L^2(\R\times S^1, R(x)
\sqrt{1+R'(x)^2}\, dx\, d\theta)$. Due to the rotational symmetry
it has a simple partial-wave decomposition; its $\,H_m\,$
component is obtained replacing $-i\partial_{\theta}$ by its
eigenvalue $m$. In this way the spectral analysis of $H$ is
reduced to a family of one-dimensional Sturm-Liouville problems.
Also the magnetic term has a natural meaning: we have
\be \label{vector potential}
 A(x)R(x) = {BR(x)^2\over 2} = {\Phi(x) \over 2\pi} = \phi(x)\,,
\ee
where $\,\phi\,$ is the magnetic flux value measured in the
standard units $\,(2\pi)^{-1}\,$, or the number of flux quanta
passing through the cross section of the cylinder.

It may be convenient to get rid of the weight factor replacing by
an operator $\tilde H$ on $L^2(\R)\otimes L^2(S^1)$. This is
achieved by the unitary transformation $\psi\mapsto R^{1/2}
(1+R'^2)^{1/4}\psi$. The only term in (\ref{Hamiltonian1}) which
changes at that is the first one: by a straightforward computation
we find
\be \label{Hamiltonian2}
 \tilde H = -\,\partial_x\, {1 \over 2(1+R'(x)^2)}\,
 \partial_x\,+\, {1\over 2R(x)^2}\, \left( -i\partial_{\theta}\,+\,
 {BR(x)^2\over 2} \right)^2 + V_{21}(x) + V_{22}(x)
\ee
with
\be
 V_{21}(x) = -\,
 {1\over 8R(x)^2}\, \left( 1+ {R(x)R''(x) \over (1+R'(x)^2)^{3/2}}
 \right)^2
\ee
and
\be
 V_{22}(x) = \left( -\, {R'^2\over 8R^2(1+R'^2)}\,-\, {7\over 8}\,
 {R'^2 R''^2 \over (1+R'^2)^3} \,+\, {R''+ R(R'R'''+ R''^2) \over
 4R(1+R'^2)^2} \right)(x)\,.
\ee

Spectral properties of the Hamiltonian are influenced by the
geometry of $S$. Suppose, \eg, that the latter has asymptotically
constant radius, $\lim_{|x|\to\infty} R(x)= R_0\,$. In the absence
of the magnetic field the problem is similar to that of a locally
deformed Dirichlet strip \cite{BGRS, EV} (it is simpler, however,
unless a mode-coupling perturbation is introduced). In the s-wave
part the effective potential $V_{21}$ creates a potential well
when $S$ is locally squeezed and a barrier in case of a
protrusion. For higher partial waves and non-zero magnetic field,
of course, the effective potential consists of several competing
contributions.


\setcounter{equation}{0}
\section{Quantum Propagation}

Our main interest in this paper is not so much the spectrum of the
Hamiltonian (\ref{Hamiltonian1}) but rather the way in which an
electron propagates over the surface of the cylinder. We will be
particularly interested in the limiting situation when the radius
modulation is gentle. This is conventionally described by means of
the scaling transformation $x\mapsto \eps x$ considering the
asymptotic behaviour as $\eps\to 0$. This can be considered as a
semiclassical limit since $\eps\to 0$ means that the  wave
packet size becomes ultimately much smaller than the length scale
of the radius variation.

It is clear from the preceding section that the effective
potential $V_2= V_{21}+V_{22}$ is then dominated by the first
term. Moreover, the operators (\ref{Hamiltonian1}) and
(\ref{Hamiltonian2}) coincide in the leading term, which will be
in the following the object of the investigation. We write its
action as
\be
  H(\eps)=-\frac{\eps^2}{2}\partial_x \frac{1}{1+
    \eps^2 V_1(x)}\partial_x+V_2(x,\eps)+
  \frac{1}{2R^2(x)}
 \left( -i\partial_{\theta}+\frac{BR^2(x)}{2}\right)^2
\ee
on a suitable domain of $L^2(\R)\otimes L^2(S^1)$ where $R(x)$,
$V_1(x)={R'(x)}^2$ are smooth on $\R$ and $V_2(x,\eps)$ is smooth
on $\R\times [-\eps_0,\eps_0]$, for some $\eps_0>0$. Introducing
an $R$-dependent operator $h(R)$ for $R\in\R_+^*$ by
\be
  h(R)=\frac{1}{2R^2}
 \left( -i\partial_{\theta}+\frac{BR^2}{2}\right)^2
\ee
on a suitable domain of $L^2(S^1)$, we can regard $H(\eps)$ as an
operator on $L^2(\R,L^2(S^1))$ that we write as
\be
   H(\eps)=-\frac{\eps^2}{2}\partial_x
   \frac{1}{1+ \eps^2V_1(x)}\partial_x+V_2(x,\eps)+h(R(x)).
\ee
The spectral analysis of $h(R)$ is straightforward and yields a
family of simple eigenvalues,
\be
  \sigma(h(R))=\{\lambda_n(R), n\in\Z\}=\left\{\frac{1}{2R^2}
  \left(n+\frac{BR^2}{2}\right)^2, n\in\Z \right\}\,,
\ee
with the corresponding eigenvectors
\be\label{eigen}
  \ffi_n(\theta)=\exp(i n\theta)/\sqrt{2\pi}\,,\quad n\in\Z.
\ee
Note that the eigenvalues $\lambda_n(R)$ correspond to the
classical effective potential $V(R)$ in (\ref{effpot}) with
$n\in\Z$ in place of $p_{\theta}$. For $n\neq m$ we have
\be
  \lambda_n(R)-\lambda_m(R)=\frac{(n-m)}{2}
  \left(\frac{(n+m)}{R^2}+B\right)
\ee
so that
\be
  \lambda_n(R)=\lambda_m(R) \quad \Leftrightarrow \quad n+m<0 \; \mbox
  { and }\; R=R_{n,m}=\sqrt{\frac{-(n+m)}{B}}\,.
\ee
Moreover,
\be
   \lambda_n(R_{n,m})=\,-\,\frac{B}{2}\frac{(n-m)^2}{(n+m)}>0\,.
\ee
Hence any pair of levels $(\lambda_n(R),\lambda_m(R))$ with
$n+m<0$ exhibits one and only one crossing as $R$ varies, whereas
other pairs never cross. The crossing points are well separated,
\be
 \{R_{n,m}:\; (n,m)\in \N^2,\, n+m<0 \}=
 \left\{\sqrt{\frac{k}{B}}\,:\; k\in\N^* \right\}\,,
\ee
with $\sqrt{\frac{k}{B}}=R_{n,-(k+n)}$, $\,n\in\N$, and the values
of the different pairs of levels crossing at $\sqrt{\frac{k}{B}}$,
for $k$ fixed, are also well separated since
\be
  \lambda_n(R_{n,-(k+n)})=\frac{B}{2}\frac{(2n+k)}{k}\,.
\ee
We note also that $\lambda_n(R)-\lambda_{-n}(R)=Bn$.

Thus, depending on our choice of function $R(x)$, the spectrum of
$h(R(x))$ may display real or avoided crossings of an arbitrary
width. Our aim is to adapt the techniques developed in \cite{hj}
to describe the propagation of Gaussian wave packets (in the
variable $x$) through these (avoided) crossings and, in
particular, the splitting of the solution among the different
angular levels $\lambda_n(R(x))$ involved. In particular, we want
to allow an $\eps$ dependent definition of the shape of our tube;
it will then turn out that the natural scale for the phenomena we
want to describe is $\delta = \sqrt{\eps}$. We henceforth adopt
$\delta$ as our small parameter and consider smooth functions
$R(x,\delta)$ defined on $\R\times [-\delta_0,\delta_0]$. This
means, in particular, that both the function $V_1$ and the
operator $h$ will depend on both $x$ and $\delta$ in a smooth
fashion.

However, the above described model can exhibit no transitions
because of the rotational invariance due to which passages between
different levels $\lambda_n$ are forbidden. To get a nontrivial
result, we perturb therefore our system by introducing a real
valued potential $\delta W(x,\theta,\delta)$, which is smooth on
$\R \times S^1 \times [-\delta_0,\delta_0]$ and violates the
symmetry. For example, we can add a constant electric field in the
direction $\vec{d}=\sin(\alpha)\vec{e}_z+\cos(\alpha)\vec{e}_x$,
where $\alpha\not\in \Z\pi$. As a consequence, we lose
integrability of the system on the classical level, whereas in the
quantum setting transitions between the different perturbed
eigenstates become possible. By assumption, when considered
as a (bounded) operator on $L^2(S^1)$ for $(x, \delta)$ fixed, the
operator $\delta W(x,\theta,\delta)$ does not commute with $h(R(x,
\delta))$, and therefore it perturbs the spectrum
$\sigma(h(R(x,\delta))$. For the time being, let us keep the
general form $\delta W(x,\theta,\delta)$ for the perturbation and
describe the differences and similarities of the present case in
comparison with the paper \cite{hj}.

We introduce the operator  $g$ on (a suitable domain of) $L^2(\R,
L^(S^1))$ by
\be
 g(x,\delta) = h(R(x,\delta))+V_2(x,\delta)+\delta W(x,\theta,\delta)
\ee
so that the perturbed full Hamiltonian reads (with a slight abuse
of notation)
\be
   H(\delta)=\,-\,\frac{\delta^4}{2}\,\partial_x \frac{1}{1+
     \delta^4V_1(x,\delta)}\partial_x+ g(x,\delta).
\ee
Without loss of generality, we can assume that
$\int_{S^1}W(x,\theta,\delta)d\theta=0$ by modifying
$V_2(x,\delta)$ if necessary. We require the different potentials
introduced so far to be smooth so that the following regularity
hypothesis is fulfilled \\[3mm]
{\bf H0:} {\it The operator $g$ is strongly $C^{\infty}$ in
$(x,\delta)$ in $\R\times [-\delta_0, \delta_0]$. \\[3mm]}
We want to approximate the solutions to the Schr\"odinger equation
in a suitable time scale,
\be\label{schro}
 i\,\delta ^2\,\frac {\partial \psi }{\partial t}\
 =\ H(\delta )\,\psi ,
\ee
for $t$ in a finite time interval, as $\delta\ra 0$, for initial
conditions of a ``coherent state'' type, which we shall describe
in detail below.

The first difference in comparison with \cite{hj} comes from the
fact that the kinetic term gives rise to a perturbed Laplacian
\bea\label{perlap}
  -\frac{\delta^4}{2}\partial_x \frac{1}{1+
     \delta^4V_1(x,\delta)}\partial_x &\!= \!& -\frac{\delta^4}{2}\partial_x^2
  +\frac{\delta^8}{2}\partial_x \frac{V_1(x,\delta)}{1+
     \delta^4V_1(x,\delta)}\partial_x\nonumber\\
 &\equiv& -\frac{\delta^4}{2}\partial_x^2 + R(x,\partial_x,\delta)
\eea
where
\be\label{decomp}
  R(x,\partial_x,\delta)=- \delta^4\frac{V_1(x,\delta)}{1+
     \delta^4V_1(x,\delta)}\frac{(-i\delta^2\partial_x)^2}{2}-
  {\delta^4\over 2} \left( -i\delta^2\partial_x\frac{V_1(x,\delta)}
     {1+\delta^4V_1(x,\delta)}\right)(-i\delta^2\partial_x).
\ee
We assume \\ [3mm]
 {\bf H1:}
\be
  \sup_{x \in \R, |\delta|\leq\delta_0}
  \left|V_1^{(k)}(x,\delta)\right|<\infty\,, \quad k=0,1.
\ee
The factor $\delta^8$ in front of the operator $R$ makes it
possible to show that the influence of this term is negligible on
the propagation of Gaussian states, so that the approximation
given in \cite{hj} remains valid. This claim is the main result of
this section and will be made precise in Proposition~\ref{aprox}
below.

The second difference in comparison with \cite{hj} is that 
unless we have and explicitly solvable situation -- and such are
rare -- we do not know in general the exact eigenvalues and
eigenstates of the operator $g(x, \delta)$. However, the
approximation derived in \cite{hj} is constructed on the basis of
this exact knowledge.  A way out is to use an incomplete
information coming from the perturbation theory. Our second
result, Theorem \ref{57}, stated in section 5 says that it
is enough to know the first few terms in the perturbation series
in order to construct an approximation that describes the
propagation, even in presence of avoided crossings,  and that
the result is as good as the one derived in \cite{hj}.

The rest of this section is organized as follows. We proceed with
the description of the ingredients needed for our approximation,
in analogy with \cite{hj}, assuming that we know the exact
diagonal form of $g(x, \delta)$. Then prove that the modification
$R$ of the Laplacian does not affect the validity of this
approximation. The next section will be devoted to the
perturbative aspects mentioned above.

We will denote by $\mu_n(x,\delta)$ the eigenvalue of
$g(x,\delta)$ such that $\mu_n(x,\delta)-\lambda_n(R(x,\delta))\ra
0$ as $\delta\ra 0$, for $x$ such that $R(x,\delta)$ far from
$R_{n,m}$. The corresponding eigenvector will be denoted by
$\Phi_n(x,\delta)$. If $R(x,\delta)$ lies in a neighborhood of
$R_{n,m}$, we will denote by $\mu_{\A}(x,\delta) \geq
\mu_{\B}(x,\delta)$ the almost degenerate perturbed eigenvalues
with corresponding eigenvectors $\Phi_{\A}(x,\delta) $ and
$\Phi_{\B}(x,\delta) $. The reason for such a convention is that
the unperturbed eigenvalues $\lambda_n(R(x,\delta)$ may or may not
cross, are therefore the labeling of the $\mu$'s in terms of the
indices $n$ and $m$ is not straightforward. Let $Q_n(x,\delta)$ be
the one-dimensional spectral projection of $g(x,\delta)$
corresponding to $\mu_n(x,\delta)$ in the first case and
$P(x,\delta)$ be the two-dimensional spectral projection of
$g(x,\delta)$ corresponding to $\mu_{\A}(x,\delta) \geq
\mu_{\B}(x,\delta)$ in the second case.

The situation we will study is that of avoided crossings of
minimum width of order $\delta$. Without loss of generality, we
can assume the avoided crossing to occur around $x=0$. More
precisely we suppose that:
\\ [3mm]
{\bf H2:} {\it The eigenvalues $\mu_{\A}(x,\delta)$ and
$\mu_{\B}(x,\delta)$ are such that
$(\mu_{\A}-\mu_{\B})^{(-1)}\{0\}={(0,0)}$ in a neighborhood of
$(0,0)$ and $\inf_{x\in I} (\mu_{\A}(x,\delta)
-\mu_{\B}(x,\delta))=c|\delta|>0$ for $\delta\neq 0$, where $c$ is
a constant and $I$ is a small interval containing $\,0$.} \\ [3mm]
We also set
\bea
  g_{\parallel}(x,\delta) &\!= \!& g(x, \delta)P(x,\delta)\, \\
  g_{\perp}(x,\delta) &\!= \!& g(x, \delta)(\,\un -P(x,\delta))\,.
\eea
We know from \cite{h} that locally around $(0,0)$ there exists an
orthonormal basis, denoted as $\{\psi_1(x,\delta),
\psi_2(x,\delta)\}$, of $P(x,\delta)L^2(S^1)$, which is regular in
$(x,\delta)$ around $(0,0)$. It is constructed in the
standard Gram-Schmidt way: we choose an orthonormal basis
$\{\psi_1,\psi_2\}$ of $P(0,0)L^2(S^1)$ and set
\bea \phi_1(x,\delta) &\!= \!& \frac{P(x,\delta)\psi_1}
  {\|P(x,\delta)\psi_1\|} \,,\\
\phi_2(x,\delta) &\!= \!& \frac{(\un - |\phi_1(x,\delta)\ket\bra
  \phi_1(x,\delta)|)P(x,\delta)\psi_2}{\|(\un - |\phi_1(x,\delta)\ket\bra
  \phi_1(x,\delta)|)P(x,\delta)\psi_2\| }\,.
\eea
There exists a $(x,\delta)$ independent unitary transform $U$ such
that in the orthonormal basis
\be\label{basis} \psi_j(x,\delta)=U\phi_j(x,\delta), \,\,\, j=1,2,
\ee
the matrix
$g_{\parallel}(x,\delta)$ takes the form
\bea\label{afh}
  g_{\parallel}(x,\delta) &\!= \!& g_1(x,\delta)
  +\bar{V}(x,\delta)\nonumber\\
 \vspace{5pts}
  &\!= \!& \pmatrix{\beta(x,\delta)&
 \gamma(x,\delta)+i\sigma(x,\delta)\cr
           \gamma(x,\delta)-i\sigma(x,\delta) & -\beta(x,\delta)}+
           \bar{V}(x,\delta)
\eea
where $\bar{V}(x,\delta)
=\mbox{trace}(g(x,\delta)P(x,\delta))$ is a regular function of
$(x,\delta)$ around the origin and
\bea\label{locom}
  \beta (x,\delta) &\!= \!& b_1x+b_2\delta+\ode (2)\,, \\
  \gamma (x,\eps) &\!= \!& c_2\delta+\ode (2)\,, \nonumber\\
  \sigma (x,\delta) &\!= \!& \ode (2)\,, \nonumber\\
  \bar{V}(x,\delta) &\!= \!& \ode (0)\,,\nonumber
\eea
where $b_1>0,\: c_2>0,\: b_2\in \R$, and the following shorthand
is used:
\be
  \ode (m)=\ode\left((x^2+\delta^2)^{m/2}\right)\,.
\ee

In order to get rid of the $\delta$-dependence in the leading
order of $\beta(x,\delta)$ in (\ref{afh}), we introduce new
variables,
\be\label{newvar}
  {x'}=b_1x+b_2\delta\,,\quad \delta ' =c_2\delta\,,\quad
  t'=b_1^2/c_2^2 t\,.
\ee
In terms of these variables, the Schr\"odinger equation
(\ref{schro}) for
\be
  \phi({x'},t')=\psi(x({x'},\delta '),t(t'))
\ee
becomes
\be
  i\delta '^2\frac{\partial}{\partial t'}\phi({x'},t')=
-\frac{\delta '^4}{2}\partial_{x'} \frac{1}{1+
     \delta '^4V_1'(x',\delta')}\partial_{x'}
  \phi({x'},t')+\frac{c_2^4}{b_1^2}g(x({x'},\delta '),\delta(\delta
  '))\phi({x'},t')
\ee
in the limit $\delta '\ra 0$, with
\be\label{newv1}
  V_1'(x',\delta')=V_1(x({x'},\delta '),\delta(\delta '))/c_2^4\,,
\ee
\be\label{dognose1}
  g_{\parallel}(x({x'},\delta '),\delta(\delta '))
  =\pmatrix{{x'}_1 &\delta '\cr
  \delta ' & -{x'}_1}+\ode (2)
  +\bar{V}(x({x'},\delta '),\delta(\delta '))\,,
\ee
where $\bar{V}(x({x'},\delta '),\delta(\delta '))$ and
$V_1'(x',\delta ') $ are regular in $({x'},\delta ')$ around
$(0,0)$ and $\ode (2)$ refers to ${x'}$ and $\delta '$. We
introduce the fixed parameter $r={c_2^4}/{b_1^2}>0$ and {\em
henceforth drop the primes on the new variables.} We assume that
$g_1(x,\delta)$ has the form (\ref{afh}) with the following local
behavior around $x=0$ and $\delta=0\,$:
\bea\label{lobe}
  \beta (x,\delta) &\!= \!&  rx+\ode (2)\,, \\
  \gamma (x,\delta) &\!= \!&  r\delta+\ode (2)\,, \nonumber\\
  \delta (x,\delta) &\!= \!&  \ode (2)\,,  \nonumber\\
  \bar{V}(x,\delta) &\!= \!&  \ode (0)\nonumber
\eea
with $r>0$.

Let us next describe the building blocks of our Born-Oppenheimer
states.

We begin with the definition of the semiclassical ''nuclear'' wave
packets denoted as $\ffi_j(A,\,B,\,\hbar,\,a,\,\eta,\,x)$. It
comes from \cite{raise}; we have specify it for our setting where
$x\in \R$. Since \cite{raise} provides a detailed discussion of
these wave packets, we refrain from proving all their properties
here.

We assume $a\in\R$, $\eta\in\R$ and $\hbar=\delta^2>0$. Let us
stress that while the last symbol is useful when adapting the
results of \cite{raise}, it has nothing to do with the Planck's
constant. We also assume that $A$ and $B$ are non-zero complex
numbers that satisfy
 \be\label{cond2} \mbox{Re}\bar{A}\,B=\,1\,.
\ee
This condition guarantees that $\mbox{Re}\,BA^{-1}$ is positive,
since $\left(\,\mbox{Re}\,BA^{-1}\,\right)^{-1}\,=\,|A|^2$.

Our definition of $\ffi_j(A,\,B,\,\hbar,\,a,\,\eta,\,x)$ is based on the
following raising operator
\be
 {\cal A}(A,B,\hbar,a,\eta)^*
 \ =\
 \frac{1}{\sqrt{2\hbar}}\,\left[\,\overline{B}\,(x-a)
 \ -\ \overline{A}\,
 (-i\hbar\frac{\partial\phantom{x}}{\partial x}-\eta)\,\right] .
\ee
\vskip .25cm \noindent {\bf Definition:}\quad For the index $j=0$,
we define the normalized complex Gaussian wave packet (modulo the
sign of the  square root) by
\bea\nonumber
 &&\ffi_0(A,\,B,\,\hbar,\,a,\,\eta,\,x)\,=\,\pi^{-1/4}\,
 \hbar^{-1/4}\,A^{-1/2}\\[6pt] &&\qquad\qquad\times\
 \exp\left\{\,-\,B\,A^{-1}\,(x-a)^2\,/(2\hbar)\,
 +\,i\,\eta\,(x-a)\,/\hbar\,\right\} .
 \eea
Then for any positive integer $j$ we define
\be
\ffi_j(A,\,B,\,\hbar,\,a,\,\eta,\,\cdot\,)=
\frac{1}{\sqrt{j!}}\
\left(\,{\cal A}(A,B,\hbar,a,\eta)^*\right)^{j}
\ffi_0(A\,,B,\,\hbar,\,a,\,\eta,\,\cdot\,).
\ee
\vskip .25cm \noindent {\bf Remarks:}\quad 1.$\;$ For $A=B=1$,
$\hbar=1$, and $a=\eta=0$, the
$\ffi_j(A,\,B,\,\hbar,\,a,\,\eta,\,\cdot\,)$ are just the standard
harmonic-oscillator eigenstates with energies $j+1/2$.\\[7pt]
2.$\;$ For each $A$, $B$, $\hbar$, $a$, and $\eta$, the set
$\{\,\ffi_j(A,\,B,\,\hbar,\,a,\,\eta,\,\cdot\,)\,\}$ is an
orthonormal basis for $L^2(\R)$.\\[7pt]
3.$\;$ The position and momentum uncertainties of the
$\ffi_j(A,\,B,\,\hbar,\,a,\,\eta,\,\cdot\,)$ are
$\sqrt{(j+1/2)\hbar}\,|A|$ and $\sqrt{(j+1/2)\hbar}\,|B|$,
respectively.\\[7pt]
4.$\;$ When we solve approximately the Schr\"odinger equation, the
choice of the sign of the square root in the definition of
$\ffi_0(A,\,B,\,\hbar,\,a,\,\eta,\,\cdot\,)$ is determined by
continuity in $t$ after an arbitrary initial choice.\\[7pt]
5.$\;$ Defining the scaled Fourier transform to be
\be\label{four}
[\,{\cal F}_{\hbar }\Psi \,](\xi )\ =\
(2\pi \hbar )^{-1/2}\,\int_{\left. I\kern-.25em R\right. } \,\Psi (x)
\,e^{-i\,\xi \,x\,/\hbar }\,dx,
\ee
then
\be\label{form}
[\,{\cal F}_{\hbar }\,\ffi _l(A,B,\hbar ,a,\eta ,\,\cdot \,)\,](\xi )
\ =\ (-i)^{l}\,e^{-i \,\eta \,a\, /\hbar }
\,\ffi _l(B,A,\hbar ,\eta ,-a ,\xi ).
\ee
We also define
\be
  V^{{\cal A}\atop {\cal B}}
 (x,\delta)=\bar{V}(x,\delta)\pm\sqrt{\beta^2(x,\delta)+
  \gamma^2(x,\delta)+\sigma^2(x,\delta)}\,,
\ee
where $x\in\R$, $\delta>0$. Let $a^{\cal C}(t)$ and $\eta^{\cal
C}(t)$ be the solutions of the classical equations of motion
\bea\label{cleq}
  \frac{d}{dt}{a^{\cal C}}(t) &\!= \!&  \eta^{\cal C}(t)\,, \\
  \frac{d}{dt}{\eta^{\cal C}}(t) &\!= \!&  -\nabla V^{\cal C}
  (a^{\cal C}(t),\delta )\,,\quad \;\;{\cal C}={\cal A},\; {\cal B}\,, \\
   \frac{d}{dt}S^{\cal C}(t) &\!= \!&  {\eta^{\cal C}}(t)^2/2- V^{\cal C}(a^{\cal C}(
t),\delta
   )\,,
\nonumber
\eea
with initial conditions
\bea\label{inco}
  a^{\cal C}(0) &\!= \!& 0\,, \\
  \eta^{\cal C}(0) &\!= \!& \eta^0(\delta)\,,\nonumber
\eea
where
\bea
  \eta^0(\delta) &\!= \!& \eta^0+\ode (\delta)\,,\quad \eta^0>0\,,\\
  S^{\cal C}(0) &\!= \!& 0\,.\nonumber
\eea
The error term $\ode (\delta )$ depends here on whether $\cal C$
is $\cal A$ or $\cal B$. In case of isolated eigenvalue $\mu_n$,
$V^{\cal C}=V^n=\mu_n$.

We further introduce complex numbers which are defined by means of
classical quantities. Let $A^{\cal C}(t)$ and $B^{\cal C}(t)$ be
the solutions of the linear system
\bea\label{cleqma}
   \frac{d}{dt}{A^{\cal C}}(t) &\!= \!& iB^{\cal C}(t)\\
    \frac{d}{dt}{B^{\cal C}}(t) &\!= \!& i{V^{\cal C}}^{(2)}(a^{\cal C}(t),\delta)A^{
\cal
    C}(t) \nonumber
\eea
where $a^{\cal C}(t)$ is the solution of (\ref{cleq}) and (\ref{inco}),
with initial conditions
\be\label{incoma}
  \matrix{A^{\cal C}(0)=A_0\,, \cr B^{\cal C}(0)=B_0\,.}
\ee
It is easy to see that these quantities actually describe the
linearized classical flow around the trajectory $(a^{\cal
C}(t),\eta^{\cal C}(t)$. The above convention regarding ${\cal C}$
applies if $\mu_n$ is isolated in the spectrum. The asymptotics of
these classical quantities for small $t$ and $\delta$ are
described in detail in Section 2 of \cite{hj}.

The determination of the ``electronic'' part of the
Born-Oppenheimer wave packet (B-O states, for short) is as
follows. Although the ``electronic'' Hamiltonian is independent of
time, it is convenient, since we deal with the time dependent
Schr\"odinger equation, to choose specific time dependent
``electronic'' eigenvectors. Since they may become singular when the
corresponding eigenvalues are degenerate, or almost degenerate, we
shall define them for $t$ in the outer regime, that is when $a(t)$
is far enough from $0$. This outer regime is characterized by
times $t$ such that (see \cite{hj})
\be\label{outertime}
  \delta^{1-\xi}\leq |t|\leq T\,, \quad \xi < 1/3\,.
\ee
We shall have two sets of eigenvectors, denoted by $\Phi_{\cal
C}^{\pm}(x,t,\delta)$, where the label $\pm$ refers to positive
and negative times. Of course, this distinction is irrelevant if
we consider an isolated eigenvalue $\mu$.

Let $\eta^{\cal C}(t)$ be the momentum solution of the classical
equations of motion(\ref{cleq}) and (\ref{inco}). The normalized
eigenvectors $\Phi_{\cal C}^{\pm}(x,t,\delta)$ are the solutions
of
\be\label{trapa}
  \bra\Phi_{\cal C}^{\pm}(x,t,\delta)|\left(\partial /\partial t+
  \eta^{\cal C}(t)\partial_x\right)\Phi_{\cal
C}^{\pm}(x,t,\delta)\ket\equiv 0
\ee
for ${\cal C}={\cal A}, {\cal B}$ and $t{>\atop <}0$. Since the
eigenvalues $\mu_{\cal A}(x,\delta)$ and $\mu_{\cal B}(x,\delta)$
are non-degenerate for any time $t$ small enough, such vectors
exist, are unique up to an overall time independent phase factors,
and are eigenvectors of $g_1(x,\delta)$ associated with $E_{\cal
C}(x,\delta)$ for any time. More precisely, we define the angles
$\ffi (x,\delta)$ and $\theta (x,\delta)$ by
\bea
  \beta( x,\delta) &\!= \!& \sqrt{\beta^2( x,\delta)+\gamma^2(
x,\delta)+\sigma^2( x,\delta)}\cos (\theta(x,\delta))\label{dog1}\\
  \gamma( x,\delta) &\!= \!& \sqrt{\beta^2( x,\delta)+\gamma^2(
x,\delta)+\sigma^2( x,\delta)}\sin(\theta(x,\delta))
  \cos (\ffi(x,\delta))\label{dog2}\\
  \sigma( x,\delta) &\!= \!& \sqrt{\beta^2( x,\delta)+\gamma^2(
x,\delta)+\sigma^2( x,\delta)}\sin(\theta(x,\delta))
  \sin (\ffi(x,\delta))\label{dog3}.
\eea
and construct static eigenvectors. Let
\bea
  \Phi_{\cal A}^-(x,\delta) &\!= \!& \e ^{i\ffi (x,\delta)}\cos
(\theta(x,\delta)/2)\psi_1(x,\delta)
  +\sin(\theta(x,\delta)/2)\psi_2(x,\delta)\\
  \Phi_{\cal B}^-(x,\delta) &\!= \!& \e ^{-i\ffi (x,\delta)}\cos
(\theta(x,\delta)/2)\psi_2(x,\delta)
  -\sin(\theta(x,\delta)/2)\psi_1(x,\delta)
\eea
be the eigenvectors of $g_1(x,\delta)$ associated with $\mu_{\cal
C}(x,\delta)$, ${\cal C}={\cal A},{\cal B}$, for $\pi/2 <
\theta(x,\delta)\leq \pi$, and
\bea
  \Phi_{\cal A}^+(x,\delta) &\!= \!& \cos (\theta(x,\delta)/2)\psi_1(x,\delta)
  +\e ^{-i\ffi (x,\delta)}\sin(\theta(x,\delta)/2)\psi_2(x,\delta)\\
  \Phi_{\cal B}^+(x,\delta) &\!= \!& \cos (\theta(x,\delta)/2)\psi_2(x,\delta)
  -\e ^{i\ffi (x,\delta)}\sin(\theta(x,\delta)/2)\psi_1(x,\delta)
\eea
be the eigenvectors of $g_1(x,\delta)$ for $0 \leq
\theta(x,\delta)< \pi/2$. The solutions of (\ref{trapa}) are of
the form
\be\label{tdev}
  \Phi_{\cal C}^{\pm}(x,t,\delta)=\Phi^{\pm}_{\cal
C}(x,\delta)\e^{i\lambda_{\cal C}^{\pm}(x,t,\delta)},\;\qquad \qquad
  \left\{\matrix{t>0\cr t<0}\right.
\ee
where $\lambda_{\cal C}^{\pm}(x,t,\delta)$ is a real valued function
satisfying the equation
\be\label{eqla}
  i\frac{\partial}{\partial t}\lambda_{\cal C}^{\pm}(x,t,\delta)+
  i\eta^{\cal C}(t)\partial_x\lambda_{\cal C}^{\pm}(x,t,\delta)+
  \bra\Phi_{\cal C}(x,\delta)|\eta^{\cal C}(t)\partial_x\Phi_{\cal
C}(x,\delta)\ket=0\,. \ee

We can get an expression for $\lambda_{\cal C}^{\pm}$ and its
derivatives as follows. We fix values of the indices and drop them
in the notation. We introduce the new variable
\be
  \omega \equiv x-a(t)
\ee
and the notation
\bea\label{resvar}
  & &{\lambda_r}(\omega,t,\delta)\equiv\lambda
(\omega+a(t),t,\delta)\,, \\
  & &{\Phi_r}(\omega,t,\delta)\equiv\Phi (\omega+a(t),\delta)\,.
  \eea
In terms of these new variables, equation (\ref{eqla}) for
${\lambda_r}$ reads
\be
  i\frac{\partial}{\partial t}{\lambda_r}(\omega,t,\delta)=-
\left\bra {\Phi_r}(\omega,t,\delta) \left|
\frac{\partial}{\partial t}{\Phi_r}(\omega,t,\delta)\right\ket
\right. \ee
with
\be
  \frac{\partial}{\partial t}{\Phi_r}(\omega,t,\delta)=
  \eta(t)\partial_x\Phi (\omega+a(t),\delta).
\ee
By integration we get
\be\label{eqpla}
  {\lambda_r}(\omega,t,\delta)=\int^t \eta(t')\bra
 \Phi  (\omega+a(t'),\delta)| \partial_x \Phi
  (\omega+a(t'),\delta)\ket dt'+
  {\lambda_r}_0(\omega,\delta)\,,
\ee
where we are free to set the integration constant ${\lambda_r}_0
(\omega,\delta)\equiv 0$.

\vskip .2in The ``nuclear'' wave function is localized around the
classical trajectory in the semiclassical regime. In view of the
genericity condition $\eta^0>0$, in the outer temporal region
major part of the ``nuclear'' wave function will be supported away
from the neighborhood where the levels almost cross. Hence we can
introduce a cutoff function which does not significantly alter the
solution and forces the support of the wave function to be away of
this neighborhood.  Let $F$ be a $C^{\infty}$
cutoff function,
\be
  F:\R^+\to\R \,,
\ee
such that
\be
  \left\{\matrix{F(r)=1 & \quad\dots\quad & 0\leq r\leq 1\cr
          F(r)=0 & \quad\dots\quad & r\geq 2}\right.
\ee
The wave functions we construct below in the outer regime will be
multiplied by the regularizing factor
\be
  F(\| x-a^{\cal C}(t)\|/\delta^{1-\delta'})\,,
\ee
where $0<\delta'<\xi$, for ${\cal C}={\cal A},{\cal B}$.

\vskip .1in \noindent{\bf Remark:} On the support of $F$ the
relation
 \be\label{xots}
  x=\eta^0(\delta)t+\ode (\delta^{1-\delta'}+t^2)
\ee
holds true, and since $\eta^0(\delta)=\eta^0+\ode (\delta)$, where
$\eta^0>0$, we find that
\be\label{kesu}
  |x|>c|t|,
\ee
uniformly in $\delta$.

A Born-Oppenheimer state $\psi_j^{{\cal C} \pm}(x,\delta,t)$ is defined by
\bea\label{bostate}
  & &\psi_j^{{\cal C}\pm}(x,\delta,t)= \\
& &F(\| x-a^{\cal C}(t)\|/\delta^{1-\delta'})
  \ffi_j(A^{\cal C}(t),\,B^{\cal C}(t),\,\delta^2,\,a^{\cal C}(t),\,
  \eta^{\cal C}(t),\,x\,)\e^{iS^{\cal C}(t)/\delta^2}
  \Phi_{\cal C}^{\pm}(x,t,\delta).\nonumber
\eea
It is a good approximation to the solution of the Schr\"odinger
equation (\ref{schro}) as $\delta\ra 0$, when $R$ is absent and
far enough of the crossing region, i.e. in the outer time regime
(\ref{outertime}), as shown in \cite{hj}. Proposition \ref{aprox}
below shows this is still true when $R$ is present.

In the inner time regime, characterized by the inequality (see \cite{hj})
\be
 -\delta^{1-\xi}\leq t\leq  \delta^{1-\xi}\,, \quad \xi<1/3\,,
\ee
we look for an approximation constructed by means of the classical
quantities associated with the potential $\bar{V}(x,\delta)$, the
average of $\mu^{\cal A}(x,\delta)$ and $\mu^{\cal B}(x,\delta)$.
Let $a(t)$ and $S(t)$ be the corresponding classical quantities
satisfying the initial conditions
\be
  \matrix{a(0)=0\,,\cr
          \eta(0)=\eta^0\,,\cr
          S(0)=0\,.}
\ee
It is suitable to use the rescaled variables
\bea\label{resval}
  \left\{\matrix{y=(x-a(t))/{\delta} \cr s=t/\delta}\right.
\eea
It is shown in \cite{hj} that a good approximation $\psi_I$ of solutions to
(\ref{schro}) in that regime, when $R$ is absent, is given by
\be\label{apin}
  \psi_I(y,s,\delta)= F(\| y\| \delta^{\delta '})\exp\left(i\frac{S(\delta
s)}{\delta^2}+
  i\frac{\eta(\delta s)y}{\delta}\right)\chi(y,s,\delta),
\ee
with
\be
  \chi(y,s,\delta)= f_0(y,s)\psi_1(a(\delta s)+\delta y,\delta)
  +g_0(y,s)\psi_2(a(\delta s)+\delta y,\delta)\,,
\ee
where $f_0,g_0$  are complex-valued functions solutions to
\be\label{lzeq}
  i\frac{\partial}{\partial s}\pmatrix{f_0(y,s)\cr g_0(y,s)}=r
  \pmatrix{\eta^0s+y&1\cr
           1&-(\eta^0s+y)}\pmatrix{f_0(y,s)\cr g_0(y,s)}\,.
\ee
The general solution to this equation is
\bea\label{pacy}
  \pmatrix{f_0(y,s)\cr g_0(y,s)} &\!=\!&
  C_1(y)\pmatrix{ \frac{(1-i)}{2}\sqrt{\frac{r}{\eta^0}}
  D_{\frac{ir}{2\eta^0}-1}\left(
 (-1+i)\sqrt{\frac{r}{\eta^0}}(\eta^0s+y)\right)\cr
  D_{\frac{ir}{2\eta^0}}\left(
  (-1+i)\sqrt{\frac{r}{\eta^0}}(\eta^0s+y)\right)}
 \\ \vspace{5pt}
  &&+\ C_2(y)\pmatrix{ D_{-\frac{ir}{2\eta^0}}\left(
  -(1+i)\sqrt{\frac{r}{\eta^0}}(\eta^0s+y)\right)\cr
  -\frac{(1+i)}{2}\sqrt{\frac{r}{\eta^0}}D_{-
  \frac{ir}{2\eta^0}-1}\left(
  -(1+i)\sqrt{\frac{r}{\eta^0}}(\eta^0s+y)\right)}.
 \nonumber
\eea
The   coefficients $C_1(y)$ and $C_2(y)$  have to be
determined by matching with the incoming solutions of B-O type at
the border of the inner and outer time regimes.

In particular, assuming for definiteness that the incoming B-O
state $\psi_{OI}$ is associated with the index $l$ for the
``nuclear'' component and the ${\cal B}$ level with the initial
momentum  $\eta^{\cal B}(0)=\eta^0$, i.e. that
\be
 \psi_{OI}(x, t, \delta)=\psi_l^{{\cal B}-}(x, t, \delta)\,,\quad
-T\leq t\leq -\delta^{1-\xi}\,, \ee
we have
\be\label{c1}
  C_1(y)\equiv 0
\ee
and
\bea\label{c2}
  C_2(y)=&-&\delta^{-1/2}\ffi_l(A_0,B_0-irA_0/\eta^0,1,0,0,y)\e^{-
\frac{\pi r}
  {8\eta^0_1}}\exp\left({\frac{i r}{2\eta^0}({y}^2-2y)}
  \right)\nonumber\\
  &\times &\exp\left( i\frac{S^{\cal B}_0(\delta,-)}{\delta^2}+
  \frac{i r}{4\eta^0}(1+3\ln(2\eta^0)+\ln r -4\ln\delta)\right),
\eea
where $S^{\cal B}_0(\delta,-)$ is real and can be computed
explicitly -- see \cite{hj}.

The analysis of \cite{hj} shows that in this situation, we get an
outgoing solution given by a linear combination of B-O states,
with explicit coefficients, associated with the same index $l$ for
the``nuclear'' component but with both levels. The initial
momentum is chosen as $\eta^{\cal A}(0)= \eta^0-2r\delta/\eta^0$
for the ${\cal A}$ level and the outgoing solution $\psi_{OO}$ is
of the form
\be
  \psi_{OO}(x, t, \delta)= -\e^{-\pi r/2\eta^0_1}\psi_l^{{\cal A}+}(x, \delta, t)
+\e^{-\pi r/4\eta^0}\sqrt{\frac{\pi r}{\eta^0}}
  \frac{\e^{i\lambda(\delta)}}{\Gamma\left(1+\frac{i
  r}{2\eta^0}\right)}\psi_l^{{\cal B}+}(x, \delta, t)
\ee
provided $\delta^{1-\xi}\leq t\leq T$, where
\be\label{lamb}
 \lambda(\delta)=\pi /4+S^{\cal A}_0
(-,\delta)/\delta^2+\frac{r}{2\eta^0} \left(1+3\ln(2\eta^0)+\ln r
-4\ln \delta\right)\,. \ee
Here again, $S^{\cal A}_0(\delta,-)$ is real and can be computed
explicitly from \cite{hj}.

Moreover, the function obtained by pasting the approximations
constructed in the outer and inner temporal regions is an
approximate solution to the Schr\"odinger equation when the
perturbation $R$ of the Laplacian is absent (see \ref{perlap}).
Similar explicit formulae are valid if the ingoing state is
associated with the ${\cal A}$ level. Hence, the propagation
through avoided crossings can be iterated.

We  are going to show that the perturbation of the Laplacian
in (\ref{perlap}) does not affect the propagation of B-O states.
The general strategy is simple: we write
\bea
  H(\delta) &\!= \!& -\frac{\delta^4}{2}\Delta_x +g(x,\delta)
  +R(x,\partial_x,\delta)\nonumber\\
  &\equiv &H_0(\delta)+R(x,\partial_x,\delta)
\eea
and denote by $\Psi_l(x,t,\delta)$ the approximation given by
$\psi_{OI}, \psi_{I}, \psi_{OO}$ in their respective time domains
constructed in \cite{hj}:
\be\label{oldstuff} \Psi_l(x,t,\delta)=\left\{\matrix{\psi_{OI}(x,
t, \delta) &\quad\dots\quad& -T\leq t\leq -\delta^{1-\xi}\cr
\psi_{I}(x, t, \delta) &\quad\dots\quad& -\delta^{1-\xi}\leq t\leq
\delta^{1-\xi}\cr \psi_{OO}(x, t, \delta) &\quad\dots\quad&
\delta^{1-\xi}\leq t\leq T }\right. \ee
We define $\zeta_l$ by
\bea\label{defxi}
\xi_l(x,t,\delta)& = &i\delta^2\partial_t \Psi_l(x,t,\delta)-H(\delta)\Psi_l(x,t,\delta)
\nonumber\\
  &\!= \!& i\delta^2\partial_t \Psi_l(x,t,\delta)-H_0(\delta)\Psi_l(x,t,\delta)-
 R(x,\partial_x,\delta)\Psi_l(x,t,\delta)\nonumber\\
  &\!= \!& \zeta_l^0(x,t,\delta)+\zeta_l^1(x,t,\delta),
\eea
where $\zeta^0_l$ is the error term controlled in \cite{hj} by means of
the following abstract lemma.

\vskip .25cm \noindent
\begin{lem}\label{fundcalc}
Suppose $H(\hbar)$ is a family of self-adjoint operators labelled
by $\hbar>0$. Suppose that $\psi(t,\,\hbar)$ belongs to the domain
of $H(\hbar)$, is continuously differentiable in $t$, and solves
approximately the Schr\"odinger equation in the sense that
\be\label{xidef} i\,\hbar\,\frac{\partial\psi}{\partial
t}(t,\,\hbar)\ =\ H(\hbar)\,\psi(t,\,\hbar)\ +\
\zeta(t,\,\hbar)\,, \ee
where $\zeta(t,\,\hbar)$, satisfies
\be\label{xiest} \|\,\zeta(t,\,\hbar)\,\|\ \le \,\mu(t,\,\hbar)\,.
\ee
Then
\be\label{ultimate} \|\,\E^{-itH(\hbar)/\hbar}\,\psi (0,\,\hbar)\
-\ \psi(t,\,\hbar)\,\|\ \le\ \hbar^{-1}\
\int_0^{t}\,\mu(s,\,\hbar)\,ds \ee
holds true for $t>0$ and the analogous statement is valid for
$t<0$.
\end{lem}

Using the same lemma to estimate the norm of $\zeta_l^1$, we get
\begin{prop} \label{aprox}
Under the hypotheses (H0)-(H2), the function $\Psi_l(x,t,\delta)$
defined by (\ref{oldstuff}) is for any $T>0$ an approximation to
the solution $\psi(x,t,\delta)$ of the Schr\"odinger equation
(\ref{schro}) such that
\be
  \psi(x,t,\delta)=\Psi_l(x,t,\delta)+\ode(\delta^p)
\ee
holds in the $L^2(\R)$ sense for some $p>0$ and all $t\in [-T,T]$.
\end{prop}

\noindent The proof of this technical proposition is given in the
appendix.


\section{Propagation of Perturbed B-O States}
\setcounter{equation}{0}

Let us now turn to the second indicated step and replace the above
B-O approximation by a construction making use of a perturbative
knowledge of the exact eigenvectors and eigenvalues of the
operator $g_{\parallel}$ defined by (\ref{afh}). In particular,
this needs to be done for the quantities appearing in
(\ref{cleq}), (\ref{cleqma}) determined by means of a classical
potential given by an approximation of the spectrum of
$g_{\parallel}$. We will show that it is enough to know the second
order perturbation expansion in order to construct an
approximation of the solution that is a perturbed version of our
initial B-O states and still describes accurately the transitions
between the ``electronic'' levels.

In order to make some explicit formulae simpler and to stress the
effect of the perturbation, we will assume in this section that
both the operators $h$ and $W$ are $\delta$ independent, i.e., we
shall consider
\be
  g(x,\delta)=h(R(x))+V_2(x) +\delta W(x,\theta)\,,
\ee
where $V_2$ commutes with $h$ whereas $W$ doesn't. This means that
$h(x)$ is assumed to have a degeneracy at $x=0$  in the
considered part of its spectrum that is lifted by $W$ to the
leading order in $\delta$.  This is the generic situation we
set out to investigate when the avoided crossing results from a
weak symmetry breaking violation of a true eigenvalue crossing.
Note, however, that we are able to accommodate the general
situation considered so far, at the cost of more complicated
perturbation formulae.

Let us state a simple lemma which is at the basis of our
constructions and which says that an approximation of an
approximate solution is an approximate solution.
\begin{lem}\label{simlem}
Let $H(\delta)$ be for all $\delta\in(0,\delta_0)$ a self adjoint
operator densely defined in a Hilbert space ${\cal H}$, and let
$\psi_a(t,\delta)\in {\cal H}$, $\ffi_a(t,\delta)\in {\cal H}$  be
time dependent vectors with the following property: there exist
$c, p_1, p_2>0$ such that the relations
\be
  \|\e^{-iH(\delta)t/\delta^2}\psi_a(0,\delta)-\psi_a(t,\delta)\|\leq c\delta^{p_1}
\ee
and
\be
  \|\ffi_a(t,\delta)-\psi_a(t,\delta)\|\leq c \delta^{p_2}
\ee
hold for all $t$ from an interval $I\subset\R$ and
$0<\delta<\delta_0$. Then
\bea
  & &\left\|\e^{-iH(\delta)t/\delta^2}\ffi_a(0,\delta)-\ffi_a(t,\delta)\right\|
  \leq 3c\delta^{\min(p_1,p_2)}\,, \nonumber\\
  & &\left\|\e^{-iH(\delta)t/\delta^2}\psi_a(0,\delta)-\ffi_a(t,\delta)\right\|
  \leq 3c\delta^{\min(p_1,p_2)}\,.
\eea
\end{lem}
{\sl Proof} uses just unitarity of the evolution group and the
Cauchy-Schwarz inequality. \ep \vskip .25cm \noindent

Our approximate B-O states will require classical quantities
defined by means of an approximation $\widetilde{V}^{\cal C}$ of
the potential $V^{\cal C}$ used in (\ref{cleq}), (\ref{cleqma}).
We have to estimate the error induced by this approximation. In
order to do that, we make use of Gronwall's lemma (see
e.g. \cite{dieu}) that we recall
below.

\begin{lem}
Let $E$ be a Banach space, $U\subset E $ be open, $I$ be an
interval of $\R$ and  $f\in C^1(I\times U;E)$ be such that there
exists $K>O$ with $\sup_{(t,x)\in I\times U}\|D_2f(t,x)\|_{{\cal
L}(E)}\leq K$. Let $g: I\times U \ra E$ be continuous and such
that there exists $G>0$ with
\be
\sup_{(t,x)\in I\times U}\|g(t,x)\|\leq G\,. \ee
If $\alpha$ and $\beta$ be $C^1$ maps from $ J\ra U$ (where
$J\subseteq I$) satisfying for $t\in J$
\bea \alpha'(t) &\!= \!& f(t,\alpha(t))\,, \\ \beta'(t) &\!= \!&
f(t,\beta(t))+\eps g(t, \beta(t))\,, \eea
then
\be
  \|\alpha(t)-\beta(t)\| \leq \|\alpha(t_0)-\beta(t_0)\|\e^{K|t-t_0|}
+\eps G(\e^{K|t-t_0|}-1)/K\,. \ee
\end{lem}

For convenience let us recall here our definition (\ref{bostate})
of a Born-Oppenheimer state $\psi_j^{\cal C}(x,\delta,t)$ in the
exterior regime:
\bea
  & &\psi_j^{\cal C}(x,\delta,t)= \\
  & &F(\| x-a^{\cal C}(t)\|/\delta^{1-\delta'})
  \ffi_j(A^{\cal C}(t),\,B^{\cal C}(t),\,\delta^2,\,a^{\cal C}(t),\,
  \eta^{\cal C}(t),\,x\,)\e^{iS^{\cal C}(t)/\delta^2}
  \Phi_{\cal C}^{\pm}(x,t,\delta)\,.\nonumber
\eea
We want to compare $\psi_j^{\cal C}(x,\delta,t)$ with an altered but
similar definition based on approximate quantum and classical
quantities for $\widetilde{\psi}_j^{\cal C}(x,\delta,t)$:
\bea\label{appbo}
  & &\widetilde{\psi_j}^{\cal C}(x,\delta,t)= \\
& &{F}(\| x-
\widetilde{a}^{\cal C}(t)\|/\delta^{1-\delta'})
  \ffi_j(\widetilde{A}^{\cal C}(t),\,\widetilde{B}^{\cal
    C}(t),\,\delta^2,
 \,\widetilde{a}^{\cal C}(t),\,
  \widetilde{\eta}^{\cal C}(t),\,x\,)\e^{i\widetilde{S}^{\cal C}(t)/\delta^2}
  \widetilde{\Phi}_{\cal C}^{\pm}(x,t,\delta)\,.\nonumber
\eea
All ``tilded'' classical quantities are generated by equations
(\ref{cleq}, \ref{cleqma}) with an approximate potential
$\widetilde{V}^{\cal C}(x,\delta)$ in place of $V^{\cal
C}(x,\delta)$. The vector
\be\label{nonstat}
 \widetilde{\Phi}_{\cal C}^{\pm}(x,t,\delta)=
 \widetilde{\Phi}_{\cal C}^{\pm}(x,\delta)\e^{
 i\widetilde{\lambda}_{\cal C}^{\pm}(x,t,\delta)}
\ee
depends on the approximate classical quantities through the phase
$\widetilde{\lambda}_{\cal C}^{\pm}$ and on an approximate
normalized quantum eigenstate $\widetilde{\Phi}_{\cal
C}^{\pm}(x,\delta)$. Note that we keep the same Gaussian function
$\ffi_j$ to construct the ``nuclear'' wave packet.

Our next goal is to apply Lemma \ref{simlem} to estimate the
errors in terms of the difference between $\widetilde{V}^{\cal C}$
and $V^{\cal C}$.
\begin{lem}\label{outeslem}
The following inequality holds in the outer time regime for the
$L^2(\R)$ norm:
\bea
  & &\|\psi_j^{{\cal C}\pm}(x,\delta,t)-\widetilde{\psi_j}^{{\cal
    C}\pm}(x,\delta,t)\| \leq \\
  & &c\left( |\widetilde{A}(t)-A(t)|+| \widetilde{B}(t)-B(t)|+
 |\widetilde{a}(t)-a(t)|/\delta^2
 +|\widetilde{\eta}(t)-\eta(t)|/\delta^2
 \right.\nonumber\\
 & & + \frac{|t|}{\delta^2}\sup_{s\in [ 0,t ]}(|\widetilde{\eta}(s)-\eta(s)|+
|\widetilde{V}({a}(s))-V(a(s))|+\sup_{x\in [\widetilde{a}(s),a(s)]}
|\partial_x\widetilde{V}(x)||\widetilde{a}(s)-a(s)|)\nonumber\\
 & &\left.+ \sup_{(x,t,\delta)}
 |F(\| x-\widetilde{a}(t)\|/\delta^{1-\delta'})
\widetilde{\Phi}_{\cal C}^{\pm}(x,t,\delta)-F(\| x-{a}(t)\|/\delta^{1-\delta'})
{\Phi}_{\cal C}^{\pm}(x,t,\delta)|. \right) \nonumber
\eea
with some constant $c$.
\end{lem}
{\sl Proof:} The index ${\cal C}$ being fixed in this context, it
will now be omitted. Other irrelevant parameters will also be
dropped in the arguments. Note that since the function $F$ is
smooth, we can write
\be
  F(\| x-\widetilde{a}(t)\|/\delta^{1-\delta'})=
 F(\| x-{a}(t)\|/\delta^{1-\delta'})+\ode ((\widetilde{a}(t)-a(t))/
 \delta^{1-\delta'})
\ee
and that the $L^2(S^1)$-norm of the vectors ${\Phi}_{\cal
C}^{\pm}(x,t,\delta)$ equals one. Since
\be
  \widetilde{V}(\widetilde{a})-V(a)=
\widetilde{V}(\widetilde{a})-\widetilde{V}({a})+\widetilde{V}({a})-V(a)
\ee
and $\eta$ and $\widetilde{\eta}$ are uniformly bounded, we infer
\bea \lefteqn{
\widetilde{S}(t)=\int_0^t\left(\widetilde{\eta}^2(s)/2-\widetilde{V}
(\widetilde{a}(s))ds \right)= S(t)+ } \\ && \hspace{-0.8cm}
\ode\left(|t|\sup_{s\in [ 0,t
]}\left(|\widetilde{\eta}(s)-\eta(s)|+
|\widetilde{V}({a}(s))-V(a(s))|+\sup_{x\in
[\widetilde{a}(s),a(s)]}
|\partial_x\widetilde{V}(x)||\widetilde{a}(s)-a(s)|\right)\right).\nonumber
\eea
Then we compute
\bea \lefteqn{
\ffi_l(\widetilde{A},\widetilde{B},\delta^2,\widetilde{a},
\widetilde{\eta},x)- \ffi_l({A},{B},\delta^2,{a},\eta,x)} \\ && =
\e^{i\widetilde{\eta}(x-\widetilde{a})/\delta^2}
\left(\ffi_l(\widetilde{A},\widetilde{B},
\delta^2,\widetilde{a},0,x)-\ffi_l(A,B,\delta^2,a,0,x)\right)
 \nonumber\\
& & +\ffi_l(A,B,\delta^2,a,0,x)\left(\e^{i\widetilde{\eta}
((a-\widetilde{a})/\delta^2)}
-\e^{i(\eta-\widetilde{\eta})(x-{a})/\delta^2}\right)\e^{i\widetilde{\eta}
(x-{a})/\delta^2}.\nonumber \eea
From Lemma 3.1 in \cite{hj} we learn that  as $\widetilde{A}\ra A$
and  $\widetilde{B}\ra B$
  \bea
  \ffi_l(\widetilde{A},\widetilde{B},\delta^2,\widetilde{a},0,x)
   &\!= \!& \ffi_l(A,B,\delta^2,a,0,x)
\nonumber\\
    &+&\ode (|\widetilde{A}-A|+| \widetilde{B}-B|+
  |\widetilde{a}-a|/\delta)\nonumber
  \eea
holds in the $L^2(\R)$ sense, 
which takes care of the first term.
Then we note that the $L^2$ norm of the remaining term is equal to
\bea \lefteqn{ \|\ffi_l(A,B,\delta^2,a,0,x)\e^{i\widetilde{\eta}
((a-\widetilde{a})/\delta^2)}-
\ffi_l(A,B,\delta^2,a,\eta-\widetilde{\eta},x) \|} \\ & & =
\|\ffi_l(B,A,\delta^2,0,-a,x)\e^{i\widetilde{\eta}
(a-\widetilde{a})/\delta^2}-\e^{-i({\eta}-\widetilde{\eta})a/\delta^2}
\ffi_l(B,A,\delta^2,\eta-\widetilde{\eta},-a,x) \|\nonumber\\ & &
=\ode\left((\widetilde{a}-a)/\delta^2
+(\widetilde{\eta}-\eta)/\delta^2\right)\nonumber  \eea
by using Plancherel formula, the properties of the $\ffi_j$ under
Fourier transform,  $\|\ffi_j\|=1$ and the above lemma again.
Then, gathering these estimates and using the facts that $A(t)$
and $B(t)$ are uniformly bounded, we get the result. \ep \vskip
.25cm

In order to use the just proved lemma, we see that it is necessary
to approximate $V^{\cal C}$ to an error of order $o(\delta^2)$ and
to show that this induces errors of the same order in the
classical trajectory $(\tilde{a}^{\cal C}, \tilde{\eta}^{\cal C}
)$ and errors of order $o(1)$ in the linearized classical flow
$(\tilde{A}^{\cal C}, \tilde{B}^{\cal C})$. Moreover, the
corresponding eigenstates $\tilde{\Phi}^{\pm}_{\cal C} $ should be
at most at a distance $o(1)$ from $\Phi^{\pm}_{\cal C} $.

When we consider times away of the matching regime, i.e. $\tau\leq
|t|\leq T $, where $\tau$ is independent of $\delta$, it is easy
to show the following result, just by using Gronwall's lemma and
regular perturbation theory. We thus omit the proof.
\begin{lem}\label{grand}
Let the time interval $(\tau,T )$ be such that the solutions to
(\ref{cleq}), (\ref{inco}) satisfy the condition
\be
 0\notin \{a^{\cal C}(t)\,\, |\,\, \tau\leq t\leq T,\,\,
 0<\delta<\delta_0\} \equiv P\,,
\ee
where the corresponding potential
\be
V^{\cal C}(x,\delta)=\mu^{\cal C}(x,\delta)\,, \quad x\in P\,, \ee
is the nondegenerate eigenvalue of $g(x,\delta)$ corresponding to
${\Phi}_{\cal C}(x,\delta)$. Let
\be
\tilde{V}^{\cal C}(x,\delta)=\mu_0^{\cal C}(x)+\delta\mu_1^{\cal C}(x)
+\delta^2\mu_2^{\cal C}(x), \, \, x\in P,
\ee
be the second-order perturbation expansion for $\mu^{\cal
C}(x,\delta)$. We define $\tilde{a}^{\cal C}, \tilde{\eta}^{\cal
C}, \tilde{A}^{\cal C}, \tilde{B}^{\cal C}, \tilde{S}^{\cal C}$ as
above with the conditions
\bea \tilde{a}^{\cal C}(\tau)={a}^{\cal C}(\tau)+o(\delta^2)\,, &&
\tilde{\eta}^{\cal C}(\tau)={\eta}^{\cal C}(\tau)+o(\delta^2)\,,
\\ \tilde{A}^{\cal C}(\tau)={A}^{\cal C}(\tau)+o(1)\,, &&
\tilde{B}^{\cal C}(\tau)={B}^{\cal C}(\tau)+o(1)\,, \\
\tilde{S}^{\cal C}(\tau)={S}^{\cal C}(\tau)+o(\delta^2)\,, && \eea
and
\be
\tilde{\Phi}_{\cal C}(x,t,\delta)={\Phi}_{\cal C}(x,0)
\e^{i\tilde{\lambda}_{\cal C}(x,t,\delta)}\,, \ee
where $\tilde{\lambda}_{\cal C}(x,t,\delta)$ is given by
(\ref{eqla}) with ${\Phi}_{\cal C}(x,0)$ in place of
${\Phi}_{\cal C}(x,\delta)$ and
\be
\tilde{\lambda}_{\cal C}(x,\tau,\delta)={\lambda}_{\cal C}(x,\tau,
\delta)+o(1)\,. \ee
Then there exists a solution $\psi(x, t, \delta)$ to the equation
(\ref{schro}) such that
\bea \lefteqn{ \psi(x, t, \delta)} \\&& \hspace{-.6cm} ={F}(\| x-
\widetilde{a}^{\cal C}(t)\|/\delta^{1-\delta'})
  \ffi_j(\widetilde{A}^{\cal C}(t),\,\widetilde{B}^{\cal
    C}(t),\,\delta^2,
 \,\widetilde{a}^{\cal C}(t),\,
  \widetilde{\eta}^{\cal C}(t),\,x\,)\e^{i\widetilde{S}^{\cal C}(t)/\delta^2}
  \widetilde{\Phi}_{\cal C}(x,t,\delta)+o(1)\nonumber
\eea
 holds true in the $L^2$-sense and for all $\tau\leq t\leq T$.
\end{lem}

\noindent {\bf Remark:} We have the familiar explicit formulae
\bea \mu_0^{\cal C}(x) &\!= \!& \mu^{\cal C}(x,0)\,, \\
\mu_1^{\cal C}(x) &\!= \!& \bra {\Phi}_{\cal C}(x, 0) |W(x)
{\Phi}_{\cal C}(x, 0)\ket\,, \\ \mu_2^{\cal C}(x) &\!= \!& -\bra
{\Phi}_{\cal C}(x, 0) |W(x) (h(x)-\mu^{\cal
C}(x,0))_r^{-1}W(x){\Phi}_{\cal C}(x, 0)\ket \,, \eea
where the reduced resolvent is given by
\be
(h(x)-\mu^{\cal C}(x,0))_r^{-1}=\sum_{j\neq {\cal
C}}\frac{|{\Phi}_{j}(x, 0)\ket\bra {\Phi}_{j}(x, 0) |}
{(\mu^j(x,0)-\mu^{\cal C}(x,0))}\,. \ee
\vskip .25cm

The above result has to be modified for times close to the
matching regime, since in that case degenerate perturbation theory
is required to define the potential. Indeed, the approximate
potential chosen in the lemma diverges as $x\ra 0$, so that
Gronwall's lemma cannot be used as it stands. Let us find the
modified potential from the perturbation theory.

The two eigenvalues of $g(x,\delta)$ which are of interest to us,
$\mu_{\cal A}(x,\delta)$ and $\mu_{\cal B}(x,\delta)$, are given
by the spectrum of $ P(x,\delta)(h(x)+\delta W(x)+V_2(x))$ This
operator is represented in the smooth orthonormal eigenbasis
(\ref{basis}) by the matrix (\ref{afh}), which we can expand to
second order in $\delta$ for any $x$ in a neighborhood of the
origin, since the projection $P(x,\delta)$ entering the definition
of the basis (\ref{basis}) is regular. Hence we can write
\be
  g_{\parallel}(x,\delta)= \pmatrix{\beta(x,\delta)&
 \gamma(x,\delta)+i\sigma(x,\delta)\cr
           \gamma(x,\delta)-i\sigma(x,\delta) & -\beta(x,\delta)}+
\bar{V}(x,\delta)\\, \ee
where
\bea \beta(x,\delta) &\!= \!& \beta_0(x)+\delta
\beta_1(x)+\delta^2\beta_2(x)+\ode(\delta^3)\equiv
B_3(x,\delta)+\ode(\delta^3)\,, \\ \gamma(x,\delta) &\!= \!&
\gamma_0(x)+\delta
\gamma_1(x)+\delta^2\gamma_2(x)+\ode(\delta^3)\equiv
G_3(x,\delta)+\ode(\delta^3)\,, \\ \sigma(x,\delta) &\!= \!&
\sigma_0(x)+\delta
\sigma_1(x)+\delta^2\sigma_2(x)+\ode(\delta^3)\equiv
S_3(x,\delta)+\ode(\delta^3)\,, \\ \bar{V}(x,\delta) &\!= \!&
\bar{V}_0(x)+\delta\bar{V}_1(x)+\delta^2
\bar{V}_2(x)+\ode(\delta^3)\equiv V_3(x,\delta)+\ode(\delta^3)\,,
\eea
with the error $\ode(\delta^3)$ being $C^{\infty}$ in $x$, and
(see (\ref{lobe}))
\bea \beta_0(x)=rx + \ode(x^2)\,, & \beta_1(x)=\ode(x)\,, \\
\gamma_0(x)= \ode(x^2)\,, & \gamma_1(x)=r+\ode(x)\,, \\
\sigma_0(x)=\ode(x^2)\,, &\sigma_1(x)=\ode(x)\,. \eea
Let us set
\be
  s(x,\delta)=\sqrt{(B_3(x,\delta))^2+(G_3(x,\delta))^2+
(S_3(x,\delta))^2}
\ee
and define our (explicit) modified potential by
\be\label{newpot}
  \widetilde{V}^{\cal C}(x,\delta)= \pm s(x,\delta)+V_3(x,\delta)
\ee
where the sign is chosen according to the value of ${\cal C}$.
It is easy to check that by construction,
\be
V^{\cal C}(x,\delta)-\widetilde{V}^{\cal C}(x,\delta)=
\ode(\delta^3) \ee
as $x\ra 0$. As above, we employ tilde to mark the values
generated by the modified potential. We only consider the dynamics
for positive times, the other case being similar.

To define the perturbed classical trajectory, we will start
integrating Newton's equations from a positive
$t_0(\delta)=\delta^{\kappa}$, for some  $2/3<\kappa<1$, using as
initial condition the explicit asymptotic expansion given in
Corollary 2.1 of \cite{hj}:
 \begin{cor}\label{outas}
In the outer regime
  $\delta\ra 0$, $t\ra 0$, $|t|/\delta \ra \infty$ and $t^3/\delta^2 \ra 0$,
we have
\bea
    a^{{\cal A}\atop {\cal B}}(t)=&-&\partial_x
\bar{V}_3(0,\delta )\frac{t^2}{2}+\eta^0(\delta)t
    \pm\frac{r}{\eta^0(\delta)}\delta t\nonumber\\
&\mp&r\left[\frac{t^2}{2}+
\frac{\delta^2\ln|t|}{2{\left( \eta^0(\delta )\right)}^2}
+\frac{\delta^2}{4{\left( \eta^0(\delta )
\right)}^2}(1+2\ln(2\eta^0(\delta)))
-\frac{\delta^2\ln\delta}{2{\left( \eta^0(\delta )\right)}^2}
    \right]\nonumber\\
     &+&\ode (t^3) +\ode (\delta^4 /t^2)\nonumber
  \eea
  The asymptotics for $\eta^{\cal C}(t)$ in the same regime is
obtained by termwise differentiation
  of the above formulae up to errors $\ode (t^2) +\ode (\delta^4
/t^3)$.
\end{cor}
The choice of $t_0(\delta)$ ensures that
\bea\label{intilde}
  \widetilde{a}(t_0) &\!= \!& a(t_0)+o(\delta^2)\,, \\
  \widetilde{\eta}(t_0) &\!= \!& \eta(t_0)+o(\delta)\,.
\eea
Whereas the error is small enough for the position, it is not the
case for the momentum. Hence we resort to energy conservation in
order to determine the momentum with a sufficient accuracy.

Let us first note that due to the uniform boundedness of the force
induced by the potentials $V^{\cal C}$ and $\widetilde{V}^{\cal C}$,
there exist a $\tau$ small but independent of $\delta$ and constants
$0<C_1<C_2<\infty$, such that as long as $t\in [-\tau, \tau]$,
\be\label{momentum} C_1<\eta^{\cal C}(t)<C_2 \,, \ee
and similarly for $ \widetilde{\eta}^{\cal C}$.

The unperturbed energy is given by
\bea E^{\cal C}(\delta) &\!= \!& (\eta^{\cal C}(t))^2/2+V^{\cal
C}(a^{\cal C}(t),\delta) =(\eta^{\cal C}_0(\delta))^2/2+V^{\cal
C}(0,\delta)\\ &\equiv& \widetilde{E}^{\cal C}(\delta)+ V^{\cal
C}(0,\delta)- \widetilde{V}^{\cal C}(0,\delta) \nonumber \eea
where the perturbed energy $\widetilde{E}^{\cal C}(\delta)$ is
explicit. Hence
\be
  \eta^{\cal C}(t)=\sqrt{2(E^{\cal C}
(\delta)-V^{\cal C}(a^{\cal C}(t),\delta))}>0
\ee
holds for  $t\in [-\tau, \tau]$, and we can define
$\widetilde{\eta}^{\cal C}(t)$ by energy conservation so that
\bea\label{etadea}
  \widetilde{\eta}^{\cal C}(t)&:=&\sqrt{2(\widetilde{E}^{\cal C}
(\delta)-\widetilde{V}^{\cal C}(\widetilde{a}^{\cal C}(t),\delta))}\\
 &\!= \!& \eta^{\cal C}(t)+\ode\left(\widetilde{V}^{\cal C}(\widetilde{a}^{\cal
  C}(t),\delta)
-  V^{\cal C}(a^{\cal
C}(t),\delta)\right)+\ode\left(\widetilde{V}^{\cal C}(0,\delta) -
V^{\cal C}(0,\delta)\right)\nonumber\\
 &\!= \!& \eta^{\cal C}(t)+\ode\left(\sup_{x\ra 0}|\widetilde{V}^{\cal C}(x,\delta)-
V^{\cal C}(x,\delta)|+ \widetilde{a}^{\cal  C}(t)-{a}^{\cal
C}(t)\right).\nonumber \eea
Thus using formula (\ref{etadea}), we make an error in $\eta^{\cal
C}$ of the same order as the error we make in $a^{\cal C}$ and
$V^{\cal C}$.

Next we turn to the approximations $\widetilde{A}^{\cal  C}(t)$
and $\widetilde{B}^{\cal  C}(t)$. They are defined as solutions to
the system (\ref{cleqma})  with  $\widetilde{V}^{\cal  C}$
in place of  ${V}^{\cal  C}$ and initial conditions at $t=\pm t_0$,
given by
\be\label{intilma}
  \pmatrix{\widetilde{A}^{{\cal A} \atop {\cal B}}(t)\cr
\widetilde{B}^{{\cal A}
\atop {\cal B}}(t)}=
   \pmatrix{A(0)\cr B(0) \mp \sign(t) irA(0)/(\eta_0(\delta))}\,.
\ee
It remains finally to consider the perturbed eigenvectors
$\widetilde{\Phi}_{\cal  C}(x,t,\delta)$ (where we dropped the
index referring to the sign of $t$). The restrictions to the
support of $F$ mentioned in lemma \ref{outeslem} and the estimate
(\ref{momentum}) imply that if we impose the condition
\be\label{defdp}
 {1-\delta'-\kappa}>0\,,
\ee
we can write
\be
  x=a(t)+\ode(\delta^{1-\delta'})\geq ct(1+ \ode(\delta^{1-\delta'-\kappa})
 \geq c\delta^{\kappa}
\ee
for some positive constant $c$, and the same estimate is true with
$a$ replaced by $\widetilde{a}$.

Hence in the considered regime the eigenvalues $\mu^{\cal C}(x,
0)$ of $P(x,0)(h(x)+V_2(x))$ display a gap that is at least of
order $x=\ode(\delta^{\kappa})$ -- see the behaviour (\ref{lobe})
-- and we call the corresponding eigenvectors $\chi_{\cal C}(x)$.
We define our perturbed static eigenvectors by
\be
 \widetilde{\Phi}_{\cal C}(x,\delta)=\chi_{\cal C}(x)
\ee
and similarly, the phase corresponding to time dependent perturbed
eigenvectors $\widetilde{\Phi}_{\cal C}(x,t,\delta)$ -- in view
(\ref{nonstat}) -- by
\be
  \widetilde{\lambda}_{{\cal C} r}(\omega,t,\delta)=
i\int_{t_0(\delta)}^t ds\; \widetilde{\eta}(s) \bra
\widetilde{\Phi}_{\cal C}(\omega+\widetilde{a}(s),\delta)  |
\partial_x\widetilde{\Phi}_{\cal C}(\omega+\widetilde{a}(s),\delta)
\ket\,, \ee
where we used the new variables (\ref{resvar}) and (\ref{eqpla}).

The next lemma tells us that our definitions of
$(\widetilde{a}^{\cal  C}(t), \widetilde{\eta}^{\cal  C}(t))$,
  $(\widetilde{A}^{\cal  C}(t), \widetilde{B}^{\cal  C}(t))$
and $\widetilde{\Phi}_{\cal C}(x,\delta, t)$
are accurate enough for our
purpose. The proof can be found in appendix.
\begin{lem}\label{lemapa} With the definitions above, there exists
a positive $\tau$ such that for all $t\in  [t_0(\delta), \tau]$ we
have
\bea
 \widetilde{a}^{\cal  C}(t) &\!= \!& {a}^{\cal  C}(t)+o(\delta^2)\,, \\
\widetilde{\eta}^{\cal  C}(t) &\!= \!& {\eta}^{\cal
C}(t)+o(\delta^2)\,, \\ \widetilde{A}^{\cal  C}(t) &\!= \!&
{A}^{\cal C}(t)+o(1)\,, \\ \widetilde{B}^{\cal  C}(t) &\!= \!&
{B}^{\cal C}(t)+o(1)\,, \\ \widetilde{\Phi}(x,t,\delta) &\!= \!&
\Phi(x,t,\delta)+o(1)\,. \eea
\end{lem}
\vskip0.25cm

Hence, with the definitions made above, we have a perturbed B-O
state given by (\ref{appbo}) that is explicitly expressed by means
of perturbation theory in $\delta$ (modulo finding the solution of
the classical equations of motion, of course) and which yields an
approximation of the solution to the Schr\"odinger equation
(\ref{schro}) for finite time intervals as $\delta\ra 0$. In
particular, putting together our results, we get the following
statement.
\begin{thm}\label{57}
Adopt the hypotheses (H1) and (H2) and assume the behaviors
(\ref{lobe}). Suppose that  $ 2/3 < \kappa
<1$ and $\tau$ is as in the above lemma. Let
$\widetilde{\psi_j}^{\cal C \pm}(x,\delta,t)$ with $|t|\geq
\delta^{\kappa}$ be a perturbed B-O states according to
(\ref{appbo}) constructed by means of the approximate quantities
considered in lemma \ref{grand} if $|t|\geq\tau$ and in lemma
\ref{lemapa} if  $\delta^{\kappa}<|t|<\tau$, subject to the condition that all
classical quantities agree at the instants $t=\pm \tau $. Let
$\psi(x,\delta, t)$ be a solution to equation (\ref{schro}) with
$\psi(x,\delta, -T)=\widetilde{\psi_j}^{\cal B -}(x,\delta,-T)$.
Then
\be
\psi(x,\delta, t)=\widetilde{\psi_j}^{\cal B -}(x,\delta,t)+o(1),
\ee
holds as $\delta\ra 0$ for all $-T\leq t\leq - \delta^{\kappa}$,
while
\be
\psi(x,\delta, t)=-\e^{-\pi r/2\eta^0}\widetilde{\psi}_j^{{\cal
A}+} (x, \delta, t) +\e^{-\pi r/4\eta^0}\sqrt{\frac{\pi
r}{\eta^0}}
  \frac{\e^{i\lambda(\delta)}}{\Gamma\left(1+\frac{i
  r}{2\eta^0}\right)}\widetilde{\psi_j}^{{\cal B}+}(x, \delta, t)+o(1)
\ee
holds for all $\delta^{\kappa}\leq t \leq T$,
with $\lambda(\delta)$ given by (\ref{lamb}).
\end{thm}

\vskip .1in \noindent{\bf Remark:}\\
It is possible also to give an explicit approximation of the wave function
in the inner time regime, $-\delta^{\kappa}\leq t\leq \delta^{\kappa}$,
in terms of quantities coming from perturbation theory. However, 
this temporal region being so short, it is not crucial for most
applications to have a detailed  approximation there.

\vskip .5cm

\section{Appendix}
\setcounter{equation}{0}

{\bf Proof of Proposition \ref{aprox}:} It is enough to show that
the  norm of $\zeta_l^1$ in (\ref{defxi}) is small and apply to
Lemma \ref{fundcalc}.
The expression (\ref{decomp}) together with H1 show that we only
need to control the effect of $p=-i\delta^2\partial_x $ and
$p^2=(-i\delta^2\partial_x )^2 $ on $ \Psi_l(x,t,\delta)$, since
for any $\psi$ we have
\be\label{break}
  \|R(x,\partial_x,\delta)\psi\|\leq C(\delta^4\|p^2\psi\|+
  \delta^6\|p\psi\|)\,.
\ee
First consider the outer temporal region and the form
(\ref{bostate}). We know from the computations in \cite{raise}
that
\bea & &\|(p-\eta)\ffi_l(A,\,B,\,\delta^2,\,a,\,\eta,\,\cdot\,)
\|= |B|\delta\sqrt{l+1/2}\,, \\ &
&\|(p-\eta)^2\ffi_l(A,\,B,\,\delta^2,\,a,\,\eta,\,\cdot\,) \|=
|B|^2\delta^2\sqrt{(6l^2+6l+3)/4}\,. \eea
Moreover, we estimate
\bea\label{espf}
  & & \left|pF\left(\|x-a^{\cal C}(t)\|/\delta^{1-\delta'}\right)\right|\leq
  c_1\delta^{1+\delta'}\,, \\
\label{esp2f}
  & &\left|p^2F\left(\|x-a^{\cal C}(t)\|/\delta^{1-\delta'}\right)\right|\leq
  c_2\delta^{2(1+\delta')}\,,
\eea
where the constants $c_1, c_2$ depend on $F$ only. Away from the
crossing region, the ``electronic'' eigenvectors are well defined
and smooth in $(x,\delta)$. Hence we only need to consider what is
going on in the neighborhood of $x=0$ to get an upper bound on the
effect of $p$ and $p^2$ on the eigenvectors $\Phi_{\cal
C}^{\pm}(x,t,\delta)$ given by (\ref{tdev}). We drop the indices
and consider
\be
 \Phi(x,t,\delta)=\e^{i\lambda(x,t,\delta)}\Phi(x,\delta)
\ee
where  $\Phi(x,\delta)$ denote some static eigenvectors and
$\lambda(x,\delta,t)$ the corresponding real valued function
defined by (\ref{eqla}).
We compute
\bea
  \partial_x\Phi(x,t,\delta) &\!= \!& \e^{i\lambda(x,t,\delta)}[\partial_x
 \Phi(x,\delta)+(i\partial_x \lambda(x,t,\delta))\Phi(x,\delta)]\,, \\
  \partial_x^2\Phi(x,t,\delta) &\!= \!& \e^{i\lambda(x,t,\delta)}[\partial_x^2
\Phi(x,\delta)+2(i\partial_x
  \lambda(x,t,\delta))\partial_x\Phi(x,\delta)\nonumber\\
& & \{i\partial_x^2 \lambda(x,t,\delta)- (\partial_x
\lambda(x,t,\delta))^2\}\Phi(x,\delta)]\,. \eea
As $\eta, \psi_j, \partial_x\psi_j,  \partial_x^2\psi_j,
  \partial_x^3\psi_j$ are all
$\ode (0)$ as $(x,t)\ra (0,0)$ in the support of $F$, we have
\bea
  \partial_x\Phi(x,\delta) &\!= \!& \ode(\partial_x\theta(x)+\partial_x\ffi(x))+\ode(
0)\,,\\
  \partial_x^2\Phi(x,\delta) &\!= \!& \ode\left((\partial_x\theta(x))^2+
  (\partial_x\ffi(x))^2
  +\partial_x^2\theta(x)+\partial_x^2\ffi(x)\right)+\ode(0)
\eea
in the norm of the ``electronic'' Hilbert space. In the expression
(\ref{eqpla}) for $\lambda$, we first check by inspection that in
all cases
\be
  \bra \Phi | \partial_x \Phi\ket =\ode\left(\partial_x\ffi\right)+\ode(0)
\ee
(see, e.g., (3.50) in \cite{hj}) since  all functions of $\theta$
and $\phi$ are uniformly bounded and, moreover, the factor of
$\partial_x\ffi$ is a function of $\theta$ only. Hence, by a
further differentiation we get
\bea
\partial_x\bra \Phi | \partial_x \Phi\ket &\!= \!&  \ode\left(\partial_x^2\ffi
 +\partial_x\ffi+\partial_x\ffi\partial_x\theta + \partial_x\theta \right)+\ode(0)\,,
\\
\partial_x^2\bra \Phi | \partial_x \Phi\ket &\!= \!&  \ode(\partial_x^3\ffi+
 \partial_x^2\ffi\partial_x\theta + \partial_x\ffi\partial_x^2\theta +
(\partial_x\ffi)^2 +\partial_x^2\ffi\nonumber\\ &
&\partial_x\ffi\partial_x\theta+\partial_x^2\theta +
(\partial_x\theta)^2 +\partial_x\theta +\partial_x\ffi
)+\ode(0)\,. \eea
It remains to estimate $\partial_x \ffi$ and $ \partial_x \theta$.
We have
\be
  \ffi (x,\delta)=\arctan (\sigma(x,\delta) / \gamma(x,\delta))
\ee
provided $\gamma(x,\delta)$ is different from zero. Hence using
(\ref{lobe}) we get
\be
  \partial_x \ffi (x,\delta)=\frac{\gamma(x,\delta)\partial_x\sigma(x,\delta)
  -\sigma (x,\delta)\partial_x \gamma (x,\delta)}{\gamma^2(x,\delta)+
  \sigma^2(x,\delta)}\,,
\ee
so that with the help of estimates of the type
$|\gamma|/\sqrt{\gamma^2+\sigma^2}\leq 1$ we arrive at
\be
  \partial_x \ffi=\ode\left(\frac{\partial_x\gamma+\partial_x\sigma}
{\sqrt{\gamma^2+\sigma^2}}\right)\,. \ee
By similar operations we eventually obtain
\be
  \partial_x^2 \ffi
  =\ode\left(\frac{\partial_x^2\gamma+\partial_x^2
  \sigma}{\sqrt{\gamma^2+\sigma^2}}\right)+
  \ode\left(\frac{(\partial_x\gamma)^2+(\partial_x\sigma)^2+
  \partial_x\gamma\partial_x\sigma}{\gamma^2+\sigma^2}\right)\
\ee
and
\bea
  \partial_x^3 \ffi  &\!= \!&  \ode\left(\frac{\partial_x^3\gamma+\partial_x^3
  \sigma}{\sqrt{\gamma^2+\sigma^2}}\right)+
  \ode\left(\frac{\partial_x^2\gamma\partial_x\sigma+\partial_x^2
  \sigma\partial_x\gamma+\partial_x^2\gamma\partial_x\gamma
  +\partial_x^2\sigma\partial_x\sigma}{\gamma^2+\sigma^2}\right)\nonumber\\
&+&
\ode\left(\frac{(\partial_x\gamma)^2\partial_x\sigma+(\partial_x
  \sigma)^2\partial_x\gamma+(\partial_x\gamma)^3+(\partial_x\sigma)^3}
 {\sqrt{\gamma^2+\sigma^2}^{3}}\right)\,.
\eea
Assuming further that
\be\label{xval}
  \| x\|=\ode(\delta^{\kappa})\,,\quad \xi<2/3 <\kappa <1-\xi<1\,,
\ee
we get from the behaviour (\ref{lobe}) in this region
\bea
  \partial_x \ffi &\!= \!& \ode\left(\frac{1}{\delta^{1-\kappa}}\right)\,, \\
  \partial_x^2 \ffi &\!= \!& \ode\left(\frac{1}{\delta}\right)\,, \\
  \partial_x^3 \ffi &\!= \!& \ode\left(\frac{1}{\delta^{2-\kappa}}\right)\,.
\eea
Then we consider
\be
 \theta(x,\delta)=\arccos\left(\frac{\beta(x,\delta)}
{\sqrt{\beta^2(x,\delta)+\gamma^2(x,\delta)+\sigma^2(x,\delta)}}\right)\,.
\ee
By computing derivatives and estimating as above, we
easily get
\bea
\partial_x\theta &\!= \!& \ode\left(\frac{\partial_x\beta}{\beta^2+\gamma^2+\sigma^2}
\right)+\ode\left(\frac{\partial_x\gamma+\partial_x\sigma}
{\sqrt{(\beta^2+\gamma^2+\sigma^2)(\gamma^2+\sigma^2)}} \right)\,,
\\
\partial_x^2\theta &\!= \!& \ode\left(\frac{\partial_x^2\beta}
{\beta^2+\gamma^2+\sigma^2} \right)+
\ode\left(\frac{\partial_x^2\gamma+\partial_x^2\sigma}
{\sqrt{(\beta^2+\gamma^2+\sigma^2)(\gamma^2+\sigma^2)}}
\right)\nonumber\\
&+&\ode\left(\frac{\partial_x\beta(\partial_x\gamma+\partial_x\sigma)}
{(\beta^2+\gamma^2+\sigma^2)\sqrt{(\gamma^2+\sigma^2)}}\right)
+\ode\left(\frac{(\partial_x\gamma)^2+(\partial_x\sigma)^2+\partial_x\gamma
\partial_x\sigma}
{\sqrt{(\beta^2+\gamma^2+\sigma^2)}(\gamma^2+\sigma^2)}\right)\nonumber\\
&+&\ode\left(\frac{(\partial_x\beta)^2}
{\sqrt{\beta^2+\gamma^2+\sigma^2}^3} \right)\,. \eea
Using (\ref{kesu}),  $\delta^{\kappa}C\geq|t|\geq \delta^{1-\xi}$,
and (\ref{xval}), we thus find
\bea
  \partial_x \theta &\!= \!& \ode\left(\frac{1}{\delta^{2(1-\xi)}}\right)\,, \\
  \partial_x^2 \theta &\!= \!& \ode\left(\frac{1}{\delta^{3(1-\xi)}}\right)\,.
\eea
Gathering the different pieces, we obtain for the derivatives of
$\lambda$ in the regime just described
\bea \lambda(x,t,\delta) &\!= \!&
 \ode(t/\delta^{1-\kappa})=\ode\left(1/\delta^{1-2\kappa}\right)\,, \\
  \partial_x\lambda(x,t,\delta) &\!= \!& \ode\left(1/\delta^{3-2\xi-2\kappa}\right)\,
, \\
  \partial_x^2\lambda(x,t,\delta) &\!= \!&
  \ode\left(1/\delta^{4-4\xi-\kappa}\right)\,.
\eea
so that we obtain the following estimates for the derivatives of the
vector $\Phi(x,t,\delta)$
\bea
  \partial_x\Phi(x,t,\delta) &\!= \!& \ode\left(1/\delta^{2-2\xi}\right)\,, \\
  \partial_x^2\Phi(x,t,\delta) &\!= \!& \ode\left(1/\delta^{4-4\xi}\right)\,.
\eea
We are now in a position to estimate the effect of $p$ and $p^2$
on the B-O states in the outer time regime:
\bea \|p\psi_l^{\cal C}\| &\!= \!& \| (pF)\ffi_l\Phi^{\cal C}+
F(p\ffi_l)\Phi^{\cal C} + F\ffi_l(p\Phi^{\cal C})\|\nonumber\\
&\leq&c\left(\delta^{1+\delta'}+\|(p-\eta^{\cal
    C})\ffi_l\|+|\eta^{\cal C}|
  +\delta^{2}|\partial_x\Phi^{\cal C}|\right)\nonumber\\
&\leq&c(l)\left(\delta^{1+\delta'}+\delta B^{\cal C}+|\eta^{\cal C}|
  +\delta^{2\xi}\right)\,.
\eea
We have already used above the fact that $|\eta^{\cal C}|$ is
uniformly bounded as $\delta$ and $t$ go to zero, and the same is
true for $B^{\cal C}$ -- see Lemma 2.1 and Proposition 2.2 in
\cite{hj}. Finally we get in the outer temporal regime
\be
  \|p\psi_l^{\cal C}\|\leq c(l)
\ee
as $\delta\ra 0$, where $c(l)$ is some constant independent of
time. By similar manipulations we also get in the same regime
\be
  \|p^2\psi_l^{\cal C}\|\leq c(l)\,.
\ee
Note that the non-vanishing term comes only from the action of $p$
on the Gaussian state $\ffi_l$, which yields essentially
$\eta^{\cal C}$ as expected, whereas the contribution from the
derivatives of the ``electronic'' eigenvectors and cutoff function
vanish. From the definition of $R$ we get a supplementary
$\delta^4$ which more than compensates for the denominator
$\delta^2$ appearing in
(\ref{ultimate})
\be
 \frac{1}{\delta^2}\int_{\delta^{1-\xi}}^{T}
  \|R(x,\partial_x,\delta)\psi_l^{\cal C}(x,t,\delta) \|dt
\leq c(l)\delta^4.
\ee

We now need to perform the same type of analysis on the
approximate wavefunction $\psi_I(y,s,\delta)$ given by
(\ref{apin}) adopted in the inner temporal region. There we use
the variables (\ref{resval}) so that the relations
\be
  \partial_y=\delta\partial_x \quad \mbox{and}\quad p=-i\delta\partial_y
\ee
have to be employed to compute the derivatives of the different
pieces in the definition of  $\psi_I(y,s,\delta)$. In this case we
need to show that
\bea\label{ines}
  \lefteqn{\frac{1}{\delta^2}\int_{-\delta^{1-\xi}}^{\delta^{1-\xi}}
  \|R(x,\partial_x,\delta)\psi_I(y(x,t),s(t),\delta) \|dt} \nonumber \\
   &\!= \!&  \frac{1}{\delta}
  \int_{-\delta^{-\xi}}^{\delta^{-\xi}}
  \left\{\int |R(\delta y+a(s\delta),\delta\partial_y,\delta)
  \psi_I(y,s,\delta)|^2 \delta dy \right\}^{1/2}
  ds\nonumber\\
  &\leq& \frac{2}{\delta^{1+\xi}}\sup_{-\delta^{-\xi}\leq s\leq
    \delta^{-\xi}} \left\{\int |R(\delta
    y+a(s\delta),\delta\partial_y,\delta)
 \psi_I(y,s,\delta)|^2 \delta dy
  \right\}^{1/2}
\eea
is of order $\delta^p$ for some positive $p$ as $\delta\ra 0$. As
above, we denoted at that the norm in the ``electronic'' Hilbert
space by a modulus. The estimates (\ref{espf}), (\ref{esp2f})
remain valid and we have
\bea
  |p\, \e^{i\eta(s \delta)y/\delta}| &\!= \!& |\eta(s \delta)|\leq C\,, \\
   |p^2 \e^{i\eta(s \delta)y/\delta}| &\!= \!& |\eta^2(s \delta)|\leq C\,,
\eea
since $\eta(t)$ is uniformly bounded in the inner temporal regime.
Noting that $x=a(\delta s)+\delta y=\ode(\delta^{1-\xi})$, we also
get from the regularity of the orthonormal basis
$\{\psi_1(x,\delta),\psi_2(x,\delta)\}$  around $(0,0)$ that
\bea
   |p\, \psi_j(a(\delta s)+\delta y,\delta)| &\!= \!& \ode (\delta^2)\,,\\
   |p^2  \psi_j(a(\delta s)+\delta y,\delta)| &\!= \!& \ode (\delta^4)
\eea
for $j=1,2$. Finally, the functions $f_0(y,s)$ and $g_0(y,s)$
determined in (\ref{pacy}) to (\ref{c2}) and their derivatives can
be estimated using the following remark. Up to phases, these
functions are given as products of a Gaussian, a polynomial in
$y$, a parabolic cylinder function times, and a factor
$1/\delta^{1/2}$ coming from the normalization of the function
$\ffi_l$. Asymptotically, these parabolic cylinder functions,
their first and second derivatives are of order $\ode
((s+\|y\|)^0)$, $\ode((s+\|y\|))$ and $\ode ((s+\|y\|)^2)$,
respectively, where $s=\ode (\delta^{-\xi})$. Hence we can write
\bea |f_0(y,s)|&\leq&P_1(y)\e^{-y^2/2|A_0|^2}\delta^{-1/2}\,, \\
|pf_0(y,s)|&\leq&P_2(y)\e^{-y^2/2|A_0|^2}\delta^{-1/2+1-\xi}\,, \\
|p^2f_0(y,s)|&\leq&P_3(y)\e^{-y^2/2|A_0|^2}\delta^{-1/2+2-2\xi}\,,
\eea
where $A_0$ is the initial condition (\ref{incoma}) and $P_j$,
$j=1,2,3$, are polynomials in $y$ the coefficients of which are
independent of $\delta$. They depend on $l$, the index of the
chosen B-O state. Similar estimates are valid for $g_0$ in place
of $f_0$. Having
\bea
  p\psi_I=\e^{iS(\delta s)/\delta^2}\e^{i\eta(s \delta)y/\delta}[
  (pF+F\eta)(f_0\psi_1+g_0\psi_2)+F((pf_0)\psi_1+(pg_0)\psi_2
 \nonumber\\ +f_0(p\psi_1)
  +g_0(p\psi_2))]
\eea
and the above estimates we can write
\bea
  |p\psi_I(y,s,\delta)|&\leq&P_4(y)\e^{-y^2/2|A_0|^2}\delta^{-1/2}
\left(\delta^{1+\delta'}+1+\delta^2+\delta^{1-\xi} \right) \eea
with another polynomial $P_4$. Hence the right hand side of
(\ref{ines}) can be further estimated to give
\be
  \frac{1}{\delta^{1+\xi}}\sup_{-\delta^{-\xi}\leq s\leq
    \delta^{-\xi}} \left\{\int |p\psi_I(y,s,\delta)|^2 \delta dy
  \right\}^{1/2}\leq c(l)/\delta^{1+\xi}.
\ee
By similar manipulations we also get
\be
  \frac{1}{\delta^{1+\xi}}\sup_{-\delta^{-\xi}\leq s\leq
    \delta^{-\xi}} \left\{\int |p^2\psi_I(y,s,\delta)|^2 \delta dy
  \right\}^{1/2}\leq c(l)/\delta^{1+\xi}.
\ee
We note that here the leading order contribution comes from
the action of $p$ on the phase $\e^{i\eta y/\delta}$ which gives
$\eta$. The supplementary factor $\delta^4$ in (\ref{break}) yields
the final estimate
\be
  \frac{1}{\delta^2}\int_{-\delta^{1-\xi}}^{\delta^{1-\xi}}
  \|R(x,\partial_x,\delta)\psi_I(y(x,t),s(t),\delta) \|dt
  \leq c(l)\delta^{3-\xi}.
\ee
Hence the proposition holds with $p=3-\xi$. \ep \vskip0.3cm

\noindent {\bf Proof of Lemma~\ref{lemapa}:} As noted above, we
cannot directly use Gronwall's lemma as stated in the text. Hence
we need to prove that the two evolutions stay close enough to each
other between times $t_0(\delta)$ and $\tau$, where $\tau$ will be
small but independent of $\delta$ by a more refined analysis. We
consider the index ${\cal A}$ and drop it in the notation.

First, it is easy to check the following asymptotic properties as
$(x,\delta)\ra (0,0)$:
\bea
 & &s(x,\delta)-\sqrt{\beta^2(x,\delta)+\gamma^2(x,\delta)+\sigma^2(x,\delta)}
 =\ode(\delta^3)\,, \\
& &\partial_x s(x,\delta)-\partial_x\sqrt{\beta^2(x,\delta)+\gamma^2(x,\delta)+\sigma
^2(x,\delta)}
 =\ode(\frac{\delta^3}{\sqrt{x^2+\delta^2}})\,, \\
& &s(x,\delta)=r\sqrt{x^2+\delta^2}(1+\ode(x+\delta))\,, \\ &
&\partial_x s(x,\delta)=\ode(1)\label{force}\,, \\ & &\partial_x^2
s(x,\delta)=\frac{r\delta^2}{(x^2+\delta^2)^{3/2}}+\ode(1)\,.
\label{secder} \eea
We collect some preliminary observations on the solution
$\widetilde{a}(t)$ to the equation
\be
  \ddot{\widetilde{a}}(t)=-\partial_x\widetilde{V}
  (\widetilde{a}(t),\delta)
\ee
for $t\in [t_0,\tau]$ with initial condition satisfying
(\ref{intilde}). We can choose $\tau>0$ independent of $\delta$,
such that
\be
a(t)-a(t_0(\delta))\geq c_0(t-t_0(\delta))\,, \ee
for some $c_0>0$ and all $t\in [t_0(\delta),\tau]$. This implies
easily by means of (\ref{intilde}) that
\be
\widetilde{a}(t)-\widetilde{a}(t_0(\delta))\geq c_1(t-t_0(\delta))
\ee
for all $t\in [t_0(\delta),\tau ]$ with a uniform constant. Hence
we can write
\be\label{conseq}
 x^2+\delta^2|_{\theta_t}\geq c_3(\delta^{\kappa}+
(t-t_0(\delta)))^2\,, \quad \forall \theta_t\in [\widetilde{a}(t),
a(t)]\,. \ee
Consider now the identities (dropping the $\delta$ dependence in
the arguments)
\bea
 & & \ddot{\widetilde{a}}(t)-\ddot{a}(t)=
 \partial_x{V}({a}(t))-
 \partial_x\widetilde{V}(\widetilde{a}(t))=\nonumber\\
& &\partial_x\sqrt{\beta^2+\gamma^2+\sigma^2}({a}(t))+
\partial_x\bar{V}({a}(t))-\partial_xs(\widetilde{a}(t))
-\partial_xV_3(\widetilde{a}(t))=\nonumber\\
& &\partial_x(\sqrt{\beta^2+\gamma^2+\sigma^2}({a}(t))-
s(a(t)))+\partial_x(\bar{V}({a}(t))-V_3(a(t)))-\nonumber\\
& &\partial^2_xs(\theta_t)(\widetilde{a}(t)-a(t))+
\partial^2_xV_3(\theta_t)(\widetilde{a}(t)-a(t))\,,
\eea
where $\theta_t\in (\widetilde{a}(t),a(t))$.
Now the first order derivatives are of order
$\delta^3/(\delta^{\kappa}+(t-t_0(\delta))$,  whereas the second
order ones are of order $\delta^3/(\delta^{\kappa}+(t-t_0))^{3}$
-- see (\ref{secder})  and (\ref{conseq}).

Hence introducing $d(t)=\widetilde{a}(t)-a(t)$ we get an ODE of
the form
\be\label{ode}
  \ddot{d(t)}=f(d(t),t) d(t)+g(d(t),t)\,,
\ee
where we have the a priori bounds
\be
  \int_{t_0}^t|f(d(s),s)|ds\leq c\delta^2 \int_{t_0}^t
  1/(\delta^{\kappa} +(s-t_0))^{3}ds\leq c \delta^{2(1-\kappa)}\,,
\ee
and since we can assume without loss that
$\delta^k+(t-t_0(\delta))<1$,
\bea
  \int_{t_0}^t|g(d(s),s)|&\leq &\int_{t_0}^t\frac{c\delta^3}{(\delta^{\kappa}+
(s-t_0))^{3}}ds\leq c\delta^3(|\ln(\delta^{\kappa})|+
|\ln(\delta^{\kappa}+(t-t_0)|)\nonumber\\
 &\!= \!& \ode(\delta^3\ln(\delta))\,.
\eea
Equation (\ref{ode}) is equivalent to
\be\label{voleq}
  d(t)=d(t_0)+(t-t_0)\dot{d}(t_0)+\int_{t_0}^tds \int_{t_0}^sdu
  (f(d(u),u) d(u)+g(d(u),u))\,.
\ee
Let us denote
\be
  D(t)=\sup_{s\in [t_0,t]}|d(s)|\,.
\ee
We deduce from the above bounds
\bea
  & &|d(t)|\leq |d(t_0)|+(t-t_0)|\dot{d}(t_0)|+c\int_{t_0}^tds\: D(s)
\delta^{2(1-\kappa)}
  +c\delta^3|\ln(\delta)|\leq\nonumber\\
  & &c ( |d(t_0)|+|\dot{d}(t_0)| +\delta^3|\ln(\delta)|
+\int_{t_0}^tds\: D(s) \delta^{2(1-\kappa)}) \eea
and, as $D$ is not decreasing,
\bea
  D(t)&\leq& c(|d(t_0)|+|\dot{d}(t_0)| +\delta^3|\ln(\delta)|+
\int_{t_0}^tds D(s)
  \delta^{2(1-\kappa)})\nonumber\\
&\leq&c(|d(t_0)|+|\dot{d}(t_0)| +\delta^3|\ln(\delta)|+D(t)
\delta^{2(1-\kappa)})\,. \eea
Since $\delta^{2(1-\kappa)}\ra 0$, we find that
\be
   D(t)
 \leq c(|d(t_0)|+|\dot{d}(t_0)| +\delta^3|\ln(\delta)|)\,.
\ee
Plugging this into (\ref{voleq}) finally yields
\be
  d(t)=d(t_0)+(t-t_0)\dot{d}(t_0)+\ode( \delta^{2(1-\kappa)}
  (|d(t_0)|+|\dot{d}(t_0)|) +\delta^3|\ln(\delta)|).
\ee
As an immediate consequence of this result and (\ref{etadea}) we
have for any $t\in [t_0(\delta),\tau]$ with our choice of
$t_0(\delta)$ and initial conditions (\ref{intilde})
\bea & &\widetilde{a}(t)-a(t)=o(\delta^2)\,, \\ &
&\widetilde{\eta}(t)-\eta(t)=o(\delta^2)\,. \eea

Turning to $(A(t), B(t))$ and their approximations, we first note
that by \cite[p.~102]{hj} we have with our choice of $t_0(\delta)$
\be
\pmatrix{\widetilde{A}(t_0)\cr \widetilde{B}(t_0)}-
\pmatrix{A(t_0)\cr B(t_0)}=o(1)\,. \ee
Then we consider the equation (equivalent to (\ref{cleqma}) and
(\ref{incoma}))
\be
  \pmatrix{{A}(t)\cr {B}(t)}=
\pmatrix{{A}(t_0)\cr {B}(t_0)}+
\int_{t_0}^t\pmatrix{O & i \cr
i\partial_x^2V(a(t))& 0} \pmatrix{A(s)\cr B(s)}
\ee
and a similar one for the approximations with the tilded symbols
everywhere. Introducing
$$\Delta(t)=\pmatrix{\widetilde{A}(t)\cr \widetilde{B}(t)}-
\pmatrix{A(t)\cr B(t)}\,,$$
we compute
\bea
  & &\Delta(t)=\Delta(t_0)+\\
 & &\int_{t_0}^t \pmatrix{0 & 0 \cr
i\partial_x^2\widetilde{V}(\widetilde{a}(s))-\partial_x^2V(a(s))&
0} \pmatrix{\widetilde{A}(s)\cr
\widetilde{B}(s)}\,ds+\int_{t_0}^t\pmatrix{0 & i \cr
i\partial_x^2V(a(s))& 0} \Delta(s)\,ds\,.\nonumber \eea
But $\left\| \pmatrix{\widetilde{A}(t)\cr \widetilde{B}(t)}
\right\| =\ode(1)$ by \cite{hj}, $\int_{t_0}^t
\partial_x^2\widetilde{V}(\widetilde{a}(s)) ds =
\ode(\delta^{2(1-\kappa)})$ and similarly for the untilded
quantities. Hence using the same type of manipulations as above,
we deduce
\be
  \|\Delta(t)\| \leq c (\delta^{2(1-\kappa)}+ \| \Delta(t_0)\|)\,.
\ee
It follows that
\bea
  \pmatrix{\widetilde{A}(t)\cr \widetilde{B}(t)}-
\pmatrix{A(t)\cr B(t)}=o(1) \eea
holds for any $t\in [t_0(\delta),\tau]$.

In order to deal with the ``electronic'' eigenvectors we consider
the perturbation series for the resolvent $(h(x)+V_2(x)+\delta
W(x)-z)^{-1}$ when the argument $z$ runs through the circle of
radius $|\mu^{\cal A}(x, 0)- \mu^{\cal B}(x, 0)|/4$ centered at
any of the eigenvalue $\mu^{\cal C}(x, 0)$. Integration on this
circle yields the eigenprojector $Q_j(x,\delta)$, $j={\cal
  A},{\cal B}$, and the estimates
\bea
 Q_j(x,\delta) &\!= \!& Q_j(x,0)+\ode(\delta W(x)/|\mu^{\cal A}(x, 0)-
\mu^{\cal B}(x, 0)|)
 =Q_j(x,0)+\ode(\delta/x)\,, \\
 \partial_x Q_j(x,\delta) &\!= \!& \partial_x Q_j(x,0)+
 \ode(\delta /|\mu^{\cal A}(x, 0)-
\mu^{\cal B}(x, 0)|^2)=\partial_xP_j(x,0)+\ode(\delta/x^2)\,. \eea
This, in turn, yields the following estimates on the eigenvectors
$\Phi_j(x,\delta)$ of the perturbed operator $h(x)+\delta W(x)$:
\bea \Phi_j(x,\delta) &\!= \!& \chi_j(x)+\ode(\delta/x)\,, \\
\partial_x\Phi_j(x,\delta) &\!= \!& \partial_x\chi_j(x)+\ode(\delta/x^2)\,.
\eea
Now we consider one eigenvector $\chi_j(x)$ and drop the index
$j$. We note here that eq.~(3.58) in \cite{hj} shows that
\be
  \lambda_r(\omega,t,\delta)=\ode(t/\delta^{1-\kappa})\,,
\ee
so that $\lambda_r(\omega,t_0(\delta),\delta)=
\ode(\delta^{2\kappa -1})\ra 0$ with $\delta$. On the other hand,
using the fact that $ \chi(x)$ is smooth and that
$\widetilde{\eta}(t)$ is uniformly bounded on $[t_0(\delta),T]$ we
find
\bea
 & & \int_{t_0(\delta)}^t i ds
 {\eta}(s)\bra {\Phi}(\omega,s,\delta)  |
\partial_x{\Phi}(\omega+{a}(s),\delta) \ket=\nonumber\\
& &\int_{t_0(\delta)}^t i ds
 (\widetilde{\eta}(s)+o(\delta^2))\nonumber\\
& &\times \bra (\widetilde{\Phi}(\omega,s,\delta) +\ode(\delta
/(a(t)+\omega)))  |
(\partial_x\widetilde{\Phi}(\omega+\widetilde{a}(s),\delta)
+\ode(\delta /(a(s)+\omega)^2)))\ket \nonumber\\ &
&=\int_{t_0(\delta)}^t i ds
 \widetilde{\eta}(s)\bra \widetilde{\Phi}(\omega,s,\delta)  |
\partial_x\widetilde{\Phi}(\omega+\widetilde{a}(s),\delta) \ket+\nonumber\\
& &o(1)+0(\delta\ln((t+\omega)/(t_0+\omega)))+\ode(\delta
(1/(t_0(\delta)+\omega)-1/(t+\omega)))\,. \eea
Having $\omega=\ode(\delta^{1-\delta'})$ and $(\ref{defdp})$, the
error terms above can be estimated by
\be
  o(1)+\ode(\delta\ln(\delta)+\delta^{1-\kappa})
\ee
which goes to zero as $\delta\ra 0$.
It follows then that
\be
   \widetilde{\lambda}(x,t,\delta)- \lambda(x,t,\delta)=o(1)
\ee
and in turn we get
\be
  \widetilde{\Phi}(x,t,\delta)- \Phi(x,t,\delta)=o(1)\,,
\ee
which concludes the proof. \ep


\subsection*{Acknowledgment}

Both authors are grateful for the hospitality extended to them
during the visits at the partner institutes, where a part of the
work was done, P.E. in UJF Grenoble-1 and A.J. in \'{U}JF AV
\v{C}R in \v{R}e\v{z}. The research has been partially supported
by GACR under the contract 1048801.


\end{document}